\documentclass[a4paper,11pt]{article}
\pdfoutput=1 
\usepackage{jcappub} 
\usepackage[T1]{fontenc} 
\usepackage{float}
\usepackage{hyperref}
\usepackage{comment}
\usepackage{tikz}
\usetikzlibrary{decorations}
\title{\boldmath 
	Dark matter spike around bumblebee black holes
}

\author[a,b,c]{S. Capozziello,}
\author[d]{S. Zare,}
\author[e]{D. F. Mota}
\author[d,1]{and H. Hassanabadi \note{Corresponding author.}
}

\affiliation[a]{Dipartimento di Fisica ``E. Pancini", Universit\`a degli Studi di Napoli, ``Federico II" \\
Complesso Universitario Monte S. Angelo,\\
	Via Cinthia 9 Edificio G, 80126 Napoli, Italy}
\affiliation[b]{Istituto Nazionale di Fisica Nucleare (INFN),\\
	Sezione di Napoli Complesso Universitario Monte S. Angelo,\\
	Via Cinthia 9 Edificio G, 80126 Napoli, Italy}
\affiliation[c]{Scuola Superiore Meridionale, Largo San Marcellino 10, 80138 Napoli, Italy}
\affiliation[d]{Faculty of Physics, Shahrood University of Technology, Shahrood, Iran\\P.O. Box 3619995161-316}
\affiliation[e]{Institute of Theoretical Astrophysics, University of Oslo, \\P.O. Box 1029 Blindern, N-0315 Oslo, Norway}

\emailAdd{capozziello@na.infn.it}
\emailAdd{soroushzrg@gmail.com}
\emailAdd{d.f.mota@astro.uio.no}
\emailAdd{hha1349@gmail.com}

\abstract{
The effects of  dark matter spike in the vicinity of the supermassive black hole,  located at the center of  M87  (the Virgo A galaxy), are investigated within the framework of the so-called Bumblebee Gravity. Our primary aim is to determine whether the background of spontaneous Lorentz symmetry breaking has a significant effect on the horizon, ergo-region, and shadow of the Kerr bumblebee black hole in the spike region.
For this purpose, we first incorporate the dark matter distribution in a Lorentz-violating  spherically symmetric space-time as a component of the energy-momentum tensors in the Einstein field equations. This leads to  a space-time metric for a Schwarzschild  bumblebee black hole with a dark matter distribution in the spike region and beyond. Subsequently, this solution is generalized to a Kerr bumblebee black hole through the use of the Newman-Janis-Azreg-A\"inou algorithm. 
Then, according to the available observational data for the dark matter spike density and radius, and the Schwarzschild radius of the supermassive black hole in  Virgo A galaxy, we examine the shapes of  shadow and demonstrate the influence of the spin parameter $a$, the Lorentz-violating parameter $\ell$ and the corresponding dark matter halo parameters $\rho_{0}$ and $r_{0}$ on the deformation and size of the shadow.	
}


\begin{document}
\maketitle
\flushbottom


\section{Introduction}
Among the most captivating astrophysical phenomena, black holes (BHs) can be used to probe  extremely strong gravity and high-energy physics, including gravitational lensing (GL),  creation of massive jets of particles, quasi-periodic oscillations, the Hawking radiation, and the disruption of nearby orbiting stars \cite{JusufiPRD2019,NampalliwarAJ2021}. Theoretically, BHs serve as a unique platform to test a variety of predictions from theories such as modified gravity \cite{Rept,Cai}, quantum gravity, and other small or large distance corrections to General Relativity (GR).

It appears from current astrophysical observations that galaxies contain supermassive BHs (SMBHs) at their centers, with the most compelling evidence located at the center of the Virgo A galaxy M87.  Specifically, a BH of about 1500 times greater mass and 2000 times greater distance compared to Sgr $\text{A}^\ast$, the BH at the center of our Milky Way galaxy, is present at the center of the Virgo A galaxy \cite{OvgunJCAP2018,KhodadiPRD2022,PantigEPJC2022}. Observations  indicated that  parameters like mass, angular momentum, and electric charge can fully describe the features of a BH. Recent electromagnetic observations of SMBHs have successfully captured the first shadow image of the BH in M87 \cite{EventHorizon,EventHorizonL4,EventHorizonL6}. 
In realistic astrophysical scenarios, an accretion mass trend is present around any SMBH due to its ability to capture light from nearby stars or accretion discs and keep it in bound orbits. BHs can be characterized by various features, including a photon sphere radius consisting of orbiting light rays. Additionally, light rays can be categorized as unstable or stable based on whether they can fall or escape to infinity, respectively \cite{ Jusufi:2022loj, JusufiMNRA2021}.
The astronomical survey carried out by the Event Horizon Telescope (EHT) team revealed that the bright accretion disk encircling the SMBH M87 appears distorted, which is believed to be caused by the phenomenon of GL. The BH gravitational pull bends light, making the region of the accretion disk, located behind the BH, visible.

Another compelling piece of evidence supporting the existence of BHs in the Universe is the detection of gravitational waves resulting from the merging of two BHs by the LIGO/Virgo team, as reported by Abbott et al. \cite{AbbottPRX2016} in 2016.
The shadow image provides valuable insight into the geometrical structure of the event horizon and the angular velocity of the BH.
The properties of BH shadows can be used to assess both General Relativity and alternative theories of gravity \cite{Rept,PsaltisPRL020}, making it a crucial aspect to consider. 
Therefore, it is crucial to continue with theoretical efforts to calculate the forms of shadows created by BHs and BH mimickers in various theories of gravity and astrophysical settings \cite{KonoplyaPLB2019,PerlickPR2022,HallaPRD2023,SolankiPRD2022,Cunha2015PRL,WeiJCAP2013,HiokiPRD2009,ShaikhPRD2019,AfrinMNRAS2021,HouJCAP2018,TangJHEP2022,PantigJCAP2022, Addazi1,Addazi2,PantigCJP2020}.
Shadow images can provide insight into various astrophysical phenomena, including accretion matter around BHs and the distribution of dark matter (DM) in the center of galaxies. In addition to  the shadow distortion, the impact of surrounding matter has also to be taken into account \cite{HouJCAP2018,TangJHEP2022,PantigJCAP2022,NishikawaPRD2019,KavanaghPRD2020,TangCPC2021,XuJCAP2021,XuJCAP2018,Pantig2022-1}.
It is worth stressing that DM is ubiquitous: It envelopes every galaxy and even permeates  the intergalactic medium \cite{NavarroAJ1997,QuinlanAJ1995}. Observational evidence for the existence of DM \cite{BullockARAA2017,PlanckAA2015,GuoMNRAS2016} includes the mass-luminosity ratio of elliptical galaxies, spiral galaxy rotation curves, baryon acoustic oscillations, cosmic microwave background radiation, and so on. Furthermore, the Cosmological Standard Model  postulates that baryonic matter constitutes a mere $5\%$ of the total mass-energy of the Universe, with the remaining $95\%$ comprised of more or less DM $27\%$ and dark energy (DE) $68\%$. In any case, besides the astrophysical evidences, the lack of evidences for the  DM at fundamental quantum level constitutes one of the most important issues of today Physics.

As such, investigating the BH shadow in the presence of dominant dark components is of great importance. 
Since the gravitational effect of DM is generally stronger than that of DE in the vicinity of a BH, the impact of DM on the properties of  BHs may be more significant than that of DE. As a result, there is a greater interest in investigating the properties of BH shadows in the presence of DM halos \cite{HouJCAP2018}.
The study of DM has been a focus of much researches, resulting in various proposed DM models, such as the cold DM (CDM) model \cite{NavarroAJ1996}, the self-interacting DM model \cite{SpergelPRL2000}, the Bose-Einstein condensation (BEC) DM model \cite{HuPRL2000,PressPRL1990,SinPRD1994,TurnerPRD1983}, the modified Newtonian dynamic model \cite{BegemanMNRAS1991}, the extended gravity models \cite{Borka1,Borka2,Annalen} and superfluid DM model \cite{BerezhianiPRD2015,JusufiEPJC2020}. 

Astronomers are interested in studying the spatial density distribution of DM, especially in the vicinity of  SMBHs or the central part of a galaxy. While the spatial density distribution of DM on  large scale, such as the outer edges of galaxies and clusters of galaxies, has been well understood \cite{PlanckAA2015}, it is not yet clear when it is in proximity of SMBHs or the central region of a galaxy \cite{XuPRD2020}.
 Hence, the distribution of DM in the vicinity of the SMBH becomes a crucial and intriguing problem to solve. The knowledge of DM distribution near a SMBH, particularly the BH in M87 and Sgr $\text{A}^\ast$, can provide significant insights to test and constrain the predictions of GR and any potential modifications beyond GR. Furthermore, it can aid in identifying potential DM candidates, very difficult to be detected by ground-based experiments.
 
 The presence of a central BH leads to a concentration of DM particles in its strong gravitational potential, forming a spike distribution near the BH horizon, as evidenced by  some previous  results \cite{UllioPRD2001,GondoloPRL1999,SadeghianPRD2013}. Typically, the DM density experiences a significant increase by several orders of magnitude owing to the gravitational field of the BH. Thus, if the DM particles   undergo annihilation, it will result in a significant increase in the intensity of gamma-ray radiation near the BH. This, in turn, provides a valuable opportunity for detecting the DM annihilation signal \cite{XuJCAP2021}.
 
 In this perspective, it is worth stressing that, in  1999, Gondolo and Silk \cite{GondoloPRL1999} examined the DM spike (DMS)  emerging when a BH experiences adiabatic growth at the center of a DM halo that initially possesses a singular power-law cusp of the form $\rho\sim r^{-\gamma}$, where $\gamma\in (0,2)$, as resulting from some  numerical simulations \cite{NavarroAJ1996,MooreAJL1998}. Their investigation revealed that the BH growth leads to the development of a DMS $\rho\sim r^{-\gamma_{\mathrm{SP}}}$ with $\gamma_{\mathrm{SP}}\in [2.25,2.5]$, which is consistent with previous scaling outcomes reported in Ref. \cite{QuinlanAJ1995}.
 While the Gondolo and Silk model uses adiabatic and Newtonian approximations, subsequent studies have incorporated general relativistic effects on the DMS near the SMBH. For example, Sadeghian et al. \cite{SadeghianPRD2013} and Ferrer et al. \cite{FerrerPRD2017} studied the effects of  DMS on Schwarzschild and Kerr BHs, respectively.  The DMS can be modeled as a polynomial function that approximates a power-law distribution. This approach has been extensively applied to investigate various phenomena associated with DM distribution close to BHs. This power-law distribution is applicable for the space-time region far from the BH event horizon. However, when considering the DM density in the vicinity of the BH, the results differ.
 
 The impact of  DM distribution on the space-time around the BH at the center of the galaxy is a significant matter of interest. One crucial aspect of this problem is how to solve the Einstein field equation while considering the presence of DM. In recent researches by Xu et al. \cite{XuJCAP2018,XuPRD2020},  this issue has been addressed by exploring the specific scenarios of $f(r) = g(r)$ and $f(r) \neq g(r)$, where $f(r)$ and $g(r)$ are metric coefficients.

The aim of the present study is to investigate how the distribution of DM affects the space-time and the shadow of a BH in  presence of Lorentz-violating (LV) terms in an extended gravity model. Specifically, we examine the impact of LV on the horizon, ergo-region, and shadow of a modified Kerr BH in the spike region (i.e. in the region where the DM density is high), using the so-called Bumblebee Gravity model. This model extends the standard framework of GR and allows for spontaneous Lorentz symmetry breaking through a non-zero vacuum expectation value of the bumblebee vector field $B_{\mu}$ under a suitable potential \cite{KhodadiEPJC2023,Khodadiarxiv2022}. The bumblebee model is an example of a theory that exhibits Lorentz violation arising from a single vector $B_{\mu}$ that acquires a non-zero vacuum expectation value, and it  is among the simplest field theories with spontaneous Lorentz and diffeomorphism violations \cite{KosteleckySamuel,Kostelecky,Bluhm,KosteleckPotting}. In this particular case, the breaking of Lorentz symmetry occurs due to the presence of a potential whose shape allows for a minimum, thereby leading to the breakdown of $U(1)$ symmetry. 
The bumblebee formalism was originally motivated by string theory, which suggests that tensor-valued fields can acquire vacuum expectation values and lead to spontaneous Lorentz symmetry breaking \cite{KosteleckyTasson}. A recent development in this area includes the derivation of the exact solution of the Schwarzschild bumblebee BH \cite{CasanaPRD2018}.

We note that, to our knowledge, no previous study has investigated the impact of the DMS on the horizon, ergo-region, and shadow of the rotating BH M87 in an extended gravitational model known as Bumblebee Gravity. This is likely due to the complexity of the metric arising from the Xu et al. method.
A recent study (Ref. \cite{NampalliwarAJ2021}) used a different method than the Xu one to derive a new BH metric incorporating the DMS density profile and investigate the shadow radius of the BH Sgr $\text{A}^\ast$. However, the study did not consider an extended gravitational model.
Accordingly, in this framework, we find that the metric coefficients are unequal $\mathcal{A}\ne\mathcal{B}$, and the function $\mathrm{K}=\sqrt{\frac{\mathcal{B}}{\mathcal{A}}} = \left(1+\ell\right)^{-1/2} r^{2} \ne r^{2}$ reflects the shift function in $\mathrm{K}$ caused by the presence of a possible Lorentz violation. Building on the method developed by Xu et al., we explore how DM alters the space-time structure of the bumblebee BH. Our investigation is aimed to understand  effects of spontaneous Lorentz symmetry breaking on the horizon, ergo-region, and shadow of the Kerr bumblebee BH in the spike region.

The structure of the article is the following: In Sec. \ref{sec2}, we introduce the DMS and two  kinds of DM halos beyond the spike region, described by the density profile of the Cold DM (CDM) model and Thomas-Fermi (TF) model, near the BH and also derive the space-time metric of the Schwarzschild bumblebee BH surrounded by DM distribution. 
In Sec. \ref{sec3},  we develop the scenario for a Kerr bumblebee BH. In Sec. \ref{sec4} and Sec. \ref{sec5},  we examine the effect of spontaneous Lorentz symmetry breaking on the horizon, ergo-region, and shadow of the Kerr bumblebee BH in the spike region. In section \ref{Conc}, we provide a summary of our findings and present our conclusions. Throughout this paper, we will adopt natural units where the fundamental constants $G$, $c$, and $\hbar$ are set to $1$.

\section{ The Schwarzschild bumblebee black hole in dark matter distribution
	\label{sec2}}
\subsection{Dark matter profiles with spike}

\begin{figure}[h]
	\centering 
	\begin{tikzpicture}[scale=1]
		\shade [inner color=blue!400, outer color=blue!20] (0,0) circle (4cm);
		\shade [inner color=black!500, outer color=black!80] (0,0) circle (0.9cm);     
		\draw [dashed, draw=black!80!black, line width=0.5pt] (0,0) circle (2.1cm);
		\draw [<->, >=latex, black!80!black, line width=1.0pt] (0,0cm) -- (-1.2,1.7cm)
		node [midway, left, font=\footnotesize, xshift=(-0.1cm)] {$R_{\mathrm{SP}}$};
		\draw (0,1.6cm) node[right] {DMS};
		\draw (-3,2.5cm) node[right] {DM halo};
		\node[rotate=-90] at (1.2,0cm) {Event Horizon};
		\draw (-0.6,-0.5cm) node[right, color=red] {SMBH};
		\draw [thick] (0,0) circle (0.9cm);
	\end{tikzpicture}
	\caption{\label{fig:11} 
Schematic plot of the presence of a central SMBH, which causes a concentration of DM distribution in the spike region and beyond. It  is caused  by the strong gravitational potential.	
}
\end{figure}
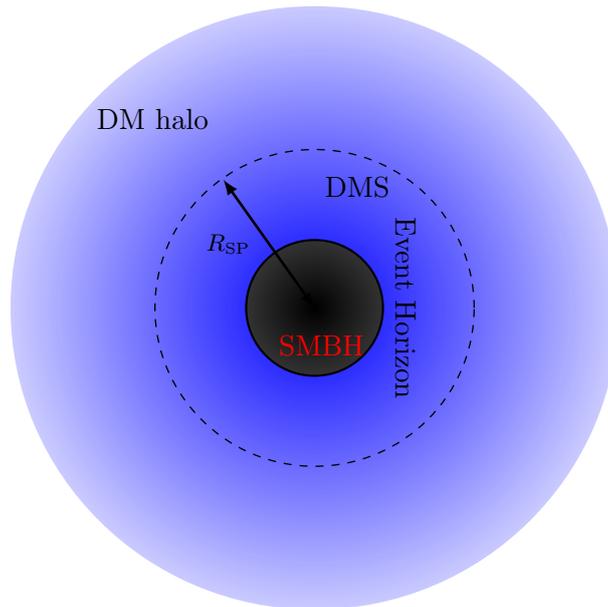

If we assume a BH  being a SMBH of mass $M_{\mathrm{BH}}$, immersed in a DM halo near a galactic center, the relevant density profile may be approximated in power-law form $\rho_{\mathrm{DM}}\simeq \rho_{0}(r_{0}/r)^{\gamma}$ with the core density $\rho_{0}$, the scale radius (or core radius) $r_{0}$ and the power-law index $\gamma$  \cite{GondoloPRL1999,NishikawaPRD2019,NampalliwarAJ2021,XuJCAP2021,KavanaghPRD2020,TangCPC2021}. 
As mentioned in Ref. \cite{GondoloPRL1999}, these assumptions give rise to  a DMS of radius $R_{\mathrm{SP}} (\gamma, M_{\mathrm{BH}})=  \mathcal{N}_{\gamma}r_{0}\left(M_{\mathrm{BH}}/\rho_{0}r_{0}^{3}\right)^{1/(3-\gamma)}$
where the normalization constant $\mathcal{N}_{\gamma}$ can be calculated numerically for 
each $\gamma$. 	The DMS  emerges from the adiabatic enhancement of the BH, which increases the central density of the host halo \cite{NampalliwarAJ2021}.
The corresponding DM density in the spike region of radius $R_{\mathrm{SP}}$ is given by $\rho_{\mathrm{R}}=\rho_{0}(R_{\mathrm{SP}}/r_{0})^{\gamma}$. 
Fig. \ref{fig:11} illustrates a schematic representation of the concentration of DM distribution in the spike region and beyond, in the vicinity of a SMBH located at the center of a galaxy. This concentration is due to the strong gravitational pull exerted by the BH.
As the DM-BH system satisfies the Newtonian approximation and the adiabatic condition, the DM density becomes	$\rho_{\mathrm{GS}}(r) = \rho_{\mathrm{R}}\left(1-4 R_{\mathrm{S}}/r\right)^{3}\left(\mathrm{R_{\mathrm{SP}}}/r\right)^{\gamma_{\mathrm{SP}}}$, where the BH Schwarzschild radius  is given by $R_{\mathrm{S}}=2M_{\mathrm{BH}} \simeq 2.95 (M_{\mathrm{BH}}/M_{\odot})$.
The value of the power-law index $\gamma_{\mathrm{SP}}$  in the DMS model, given by $\gamma_{\mathrm{SP}} = (9-2\gamma)/(4-\gamma)$, depends on the initial DM profile. 
As shown in Ref. \cite{NishikawaPRD2019}, the density profile of DMS varies from the Navarro-Frenk-White (NFW) density profile, which is fixed by numerical simulations of collisionless DM particles in galactic halos, for values of $\gamma = 1$ and $\gamma = 0$, and for SMBH masses of $M_\mathrm{BH} = 10^{5}M_{\odot}$ or $M_\mathrm{BH} = 10^{6}M_{\odot}$.
As reported in Ref. \cite{SadeghianPRD2013}, the density profile for DMS assigned for full relativistic case is 
$\rho_{\mathrm{FR}}(r) = \rho_{\mathrm{R}}\left(1-2 R_{\mathrm{S}}/r\right)^{3}\left(\mathrm{R_{\mathrm{SP}}}/r\right)^{\gamma_{\mathrm{SP}}}$. Note that $R_{\mathrm{SP}}$ is the same for both Newtonian approximation and the full relativistic case.
Therefore, the  general density profile of the DMS model around the Schwarzschild BH is given by 
\begin{equation}\label{DMSprofile}
	\rho_{\mathrm{DMS}}(r) = \rho_{\mathrm{R}}\left(1-\frac{\mathcal{K} R_{\mathrm{S}}}{r}\right)^{3}\left(\frac{\mathrm{R_{\mathrm{SP}}}}{r}\right)^{\gamma_{\mathrm{SP}}},
\end{equation}	
so that this distribution is defined in the range $\mathcal{K}R_{\mathrm{S}}<r<R_{\mathrm{SP}}$. The distance from the center of the BH is denoted by $r$. Meanwhile, $\mathcal{K}$ is a constant parameter assuming the values  4 and 2 for Newtonian approximation and full relativistic case, respectively.

To proceed with the approach, the first step to construct the metric coefficient function for a pure DM model is founding the corresponding mass distribution of the DM model. The mass profile for the DM is 
\begin{equation}\label{MassProfDMH}
	M_{\mathrm{DM}}(r) = 4\pi \int_{r_{b}}^{r} \rho_{\mathrm{DM}}(r')r'^{2}\, \mathrm{d}r'.
\end{equation}
where, the mass $M_{\mathrm{DM}}$, corresponding to density profile \eqref{DMSprofile}, is \cite{XuJCAP2021}
\begin{equation}\label{MassDMS}
	\begin{split}
		M_{\mathrm{DMS}}(r) &=  4\pi \rho_{\mathrm{R}} \left(\frac{\mathrm{R_{\mathrm{SP}}}}{r}\right)^{\gamma_{\mathrm{SP}}}\left(\frac{r^{3}-\left(\mathcal{K} R_{\mathrm{S}}\right)^{3-\gamma_{\mathrm{SP}}}r^{\gamma_{\mathrm{SP}}}}{3-\gamma_{\mathrm{SP}}}-\frac{3\mathcal{K}R_{\mathrm{S}}\left(r^{2}-\left(\mathcal{K} R_{\mathrm{S}}\right)^{2-\gamma_{\mathrm{SP}}}r^{\gamma_{\mathrm{SP}}}\right)}{2-\gamma_{\mathrm{SP}}}\right.\\
		&\left.+\frac{3\mathcal{K}^{2}R_{\mathrm{S}}^{2}\left(r-\left(\mathcal{K} R_{\mathrm{S}}\right)^{1-\gamma_{\mathrm{SP}}}r^{\gamma_{\mathrm{SP}}}\right)}{1-\gamma_{\mathrm{SP}}}+\frac{\mathcal{K}^{3}R_{\mathrm{S}}^{3}\left(1-\left(\mathcal{K} R_{\mathrm{S}}\right)^{-\gamma_{\mathrm{SP}}}r^{\gamma_{\mathrm{SP}}}\right)}{\gamma_{\mathrm{SP}}}
		\right), \quad r_{b} =\mathcal{K} R_{\mathrm{S}}.
	\end{split}
\end{equation}
Furthermore,  there is a relation between the tangential velocity of test particles with  mass $M_{\mathrm{DM}}$, that is, $\mathcal{V}_{tg}^{2}(r) = M_{\mathrm{DM}}/r$ \cite{Pantig2022-1,PantigEPJC2022}. 
Next, based on the following DM line element  given by 
\begin{equation}\label{Metric1}
	\begin{split}
		\mathrm{d}s^{2}&=\mathrm{g}_{\mu\nu} \mathrm{d}x^{\mu}\mathrm{d}x^{\nu}\\
		&=-A(r)\mathrm{d}t^{2}+B(r)^{-1}\mathrm{d}r^{2}+C(r)\left(\mathrm{d}\theta^{2}+\text{sin}^{2}\theta\mathrm{d}\varphi^{2}\right), 
	\end{split}
\end{equation}
with $A(r) = B(r)$ and $C(r) = r^{2}$,
a rotational velocity, in the equatorial plane belonging to test particle in spherical  symmetric  space-time,  can be obtained by
\begin{equation}\label{rotationalvelocity1}
	\mathcal{V}_{tg}^{2}(r) = \frac{r}{\sqrt{A(r)}}\frac{\mathrm{d}\sqrt{A(r)}}{\mathrm{d}r} = r \frac{\mathrm{d}}{\mathrm{d}r} \text{ln}\left(\sqrt{A(r)}\right).
\end{equation}
From Ref. \cite{Matos2005}, one can see that the difference in physical effect between $A(r) = B(r)$ or $A(r) \neq B(r)$  is much smaller than the effect of  DM on the BH. Therefore, Eq. \eqref{rotationalvelocity1} is valid for both cases $A(r) = B(r)$ and $A(r) \neq B(r)$.
Thus, the rotational velocity corresponding to the mass $M_{\mathrm{DMS}}$ can be calculated as
\begin{equation}\label{rotationalvelocityDMS}
	\begin{split}
		\mathcal{V}_{\mathrm{DMS}}(r) &= \left(\frac{4\pi \rho_{\mathrm{R}}}{r} \left(\frac{\mathrm{R_{\mathrm{SP}}}}{r}\right)^{\gamma_{\mathrm{SP}}}\left(\frac{r^{3}-\left(\mathcal{K} R_{\mathrm{S}}\right)^{3-\gamma_{\mathrm{SP}}}r^{\gamma_{\mathrm{SP}}}}{3-\gamma_{\mathrm{SP}}}-\frac{3\mathcal{K}R_{\mathrm{S}}\left(r^{2}-\left(\mathcal{K} R_{\mathrm{S}}\right)^{2-\gamma_{\mathrm{SP}}}r^{\gamma_{\mathrm{SP}}}\right)}{2-\gamma_{\mathrm{SP}}}\right.\right.\\
		&\left.\left.+\frac{3\mathcal{K}^{2}R_{\mathrm{S}}^{2}\left(r-\left(\mathcal{K} R_{\mathrm{S}}\right)^{1-\gamma_{\mathrm{SP}}}r^{\gamma_{\mathrm{SP}}}\right)}{1-\gamma_{\mathrm{SP}}}+\frac{\mathcal{K}^{3}R_{\mathrm{S}}^{3}\left(1-\left(\mathcal{K} R_{\mathrm{S}}\right)^{-\gamma_{\mathrm{SP}}}r^{\gamma_{\mathrm{SP}}}\right)}{\gamma_{\mathrm{SP}}}
		\right)\right)^{1/2}.
	\end{split}
\end{equation}
Subsequently, the metric coefficient function arising from the DM density profile can be determined using the rotational velocity in Eq. \eqref{rotationalvelocity1} as
\begin{equation}\label{DMmetricCoeffFuncA1}
	A(r) = \mathrm{e}^{2 \int \frac{\mathcal{V}_{tg}^{2}(r)}{r}\, \mathrm{d}r},
\end{equation}
Thereby, the function $A(r)$ associated with the density profile of the DMS is
\begin{equation}\label{DMSmetricCoeffFuncA1}
	\begin{split}
		A(r) &= r^{\frac{48\pi\rho_{\mathrm{R}}R_{\mathrm{SP}}^{\gamma_{\mathrm{SP}}}\left(\mathcal{K}R_{S}\right)^{3-\gamma_{\mathrm{SP}}}}{\gamma_{\mathrm{SP}}\left(\gamma_{\mathrm{SP}}-1\right)\left(\gamma_{\mathrm{SP}}-2\right)\left(\gamma_{\mathrm{SP}}-3\right)}} \,\,\mathrm{exp}\left(\frac{8\pi\rho_{\mathrm{R}}R_{\mathrm{SP}}^{\gamma_{\mathrm{SP}}}}{\left(\gamma_{\mathrm{SP}}-3\right)^{2}}r^{3-\gamma_{\mathrm{SP}}}-\frac{24\pi\rho_{\mathrm{R}}R_{\mathrm{SP}}^{\gamma_{\mathrm{SP}}}\mathcal{K}R_{S}}{\left(\gamma_{\mathrm{SP}}-2\right)^{2}}r^{2-\gamma_{\mathrm{SP}}}\right.\\
		&\left.+\frac{24\pi\rho_{\mathrm{R}}R_{\mathrm{SP}}^{\gamma_{\mathrm{SP}}}\left(\mathcal{K}R_{S}\right)^{2}}{\left(\gamma_{\mathrm{SP}}-1\right)^{2}}r^{1-\gamma_{\mathrm{SP}}}-\frac{8\pi\rho_{\mathrm{R}}R_{\mathrm{SP}}^{\gamma_{\mathrm{SP}}}\left(\mathcal{K}R_{S}\right)^{3}}{\gamma_{\mathrm{SP}}^{2}}r^{-\gamma_{\mathrm{SP}}}
		\right).
	\end{split}
\end{equation}	
If $R_{\mathrm{SP}}$ goes to zero, we see that the function $A(r)$ tends to one, which means that the DM space-time reduces to the Minkowski space-time.

\subsection{Dark matter halo profiles
}	
At large scales, where $R_{\mathrm{SP}}\ll r$, the DMS halo evolves into a DM halo, such that $R_{\mathrm{S}} =0$, $\gamma_{\mathrm{SP}} = \gamma$, $R_{\mathrm{SP}} = R_{c}$, and $\rho_{\mathrm{R}}=\rho_{c}$ in metric coefficient \eqref{DMSmetricCoeffFuncA1}. In the CMD model, $\gamma = 1$, while in the TF model, $\gamma = 0$ \cite{XuJCAP2018}. The space-time metric, constructed by the metric coefficient, matches with either the NFW profile associated with the CMD model or the TF profile, and describes the region away from the BH where the DM halo dominates.

\subsubsection{ The cold dark matter profile}
The NFW profile is a well-known density profile that is consistent with astronomical observations at large scale and it is derived from numerical simulations based on the $\Lambda\text{CDM}$ model, where $\Lambda$ is the cosmological constant. Although the physical nature of DM is still unknown, it is modeled as non-relativistic motion for DM particles \cite{PantigEPJC2022}. In the CDM model \cite{NavarroAJ1996,NavarroAJ1997}, the NFW profile is the corresponding distribution, which can be expressed as follows
\begin{equation}\label{CDMprof}
	\rho_{\mathrm{CDM}}(r) = \frac{\rho_{c}}{\frac{r}{R_{c}}\left(1+\frac{r}{R_{c}}\right)^{2}},
\end{equation}
where $\rho_{\mathrm{CDM}}$ is the density of the Universe at the time of the DM collapse, and $\rho_{c}$ and $R_{c}$ are the core density and core radius respectively \cite{XuJCAP2018}. 
Then, the mass $M_{\mathrm{DM}}$, corresponding to density profile \eqref{CDMprof}, is
\begin{equation}\label{MassCDM}
	M_{\mathrm{CDM}}(r) = 4\pi R_{c}^{3} \rho_{c} \left(-1+\frac{R_{c}}{r+R_{c}}-\text{ln} \left(R_{c}\right)+\text{ln}\left(r+R_{c}\right)\right).
\end{equation}
with the assumption $r_{b} =0$.
Thus, the rotational velocity corresponding to the mass $M_{\mathrm{CDM}}$ can be calculated as
\begin{equation}\label{rotationalvelocity2}
	\mathcal{V}_{\mathrm{CDM}}(r) = \sqrt{\frac{4\pi\rho_{c}R_{c}^{3}}{r}\left(\text{ln}\left(1+\frac{r}{R_{c}}\right)-\frac{r}{r+R_{c}}\right)}.
\end{equation}
Hence, the function $A(r)$ associated with the density profile of the CDM can be written as
\begin{equation}\label{DMmetricCoeffFuncA2}
	A(r) = \left(1+\frac{r}{R_{c}}\right)^{-\frac{8\pi \rho_{c} R_{c}^{3}}{r}}.
\end{equation}

\subsubsection{ The Thomas-Fermi profile}
The BEC-DM model offers an interesting DM density distribution that differs from the CDM model \cite{HuPRL2000,PressPRL1990,SinPRD1994,TurnerPRD1983}. While the behavior of DM in the BEC model is consistent with the CDM model at larger scales such as outside the galaxy, galaxy group, and the Universe, the DM density approaches a constant as $r$ approaches $0$ \cite{Bohmer2007}. In the BEC-DM model with TF approximation, the DM density profile is given by:
\begin{equation}\label{TFprof}
	\rho_{\mathrm{TF}}(r) = \rho_{c} \frac{\text{sin}(kr)}{kr}, \qquad k=\pi/R_{c},
\end{equation}
With regard to Eqs. \eqref{MassProfDMH} and  \eqref{TFprof}, the mass   is
\begin{equation}\label{MassTF}
	M_{\mathrm{TF}}(r) = \frac{4 \rho_{c} R_{c}^{2} \,r}{\pi} 
	\left(\frac{R_{c}}{\pi r} \,\text{sin}\left(\frac{\pi r}{R_{c}}\right)-\text{cos}\left(\frac{\pi r}{R_{c}}\right)\right),
\end{equation} 
with $r_{b} =0$. 
and  the rotational velocity  
$\mathcal{V}_{\mathrm{TF}}(r) = \sqrt{M_{\mathrm{TF}}/r}$
can be derived.  Then, by applying Eq. \eqref{DMmetricCoeffFuncA1}, we arrive to
\begin{equation}\label{DMmetricCoeffFuncA3}
	A(r) = \mathrm{e}^{-\frac{8\rho_{c} R_{c}^{3}}{\pi^{2}r}\text{sin}\left(\frac{\pi r}{R_{c}}\right)}.
\end{equation}


\subsection{The Lorentz-violating spherically symmetric solution in  dark matter distribution}
As it  can be seen, $A(r)$ is responsible for the DM profile information, and one of the purposes of this paper is to combine it with a spherically symmetric exact solution from the gravity sector contained in the minimal standard-model extension (SME).  The solution belongs to a theoretical model that couples a Riemann space-time with the bumblebee field responsible for the spontaneous Lorentz symmetry breaking   \cite{CasanaPRD2018}.
The resultant solution of the modified Einstein equation, generated by the bumblebee Gravity, establishes a Schwarzschild-like BH.
The space-time metric of Schwarzschild bumblebee BH is given by 
\begin{equation}\label{metricSchwlike1}
	\mathrm{d}s^{2}= -\left(1-\frac{2M_{\mathrm{BH}}}{r}\right)\mathrm{d}t^{2}+\left(1+\ell\right)\left(1-\frac{2M_{\mathrm{BH}}}{r}\right)^{-1}\mathrm{d}r^{2}+r^{2}\left(\mathrm{d}\theta^{2}+\text{sin}^{2}\theta\mathrm{d}\varphi^{2}\right).
\end{equation}
This LV spherically symmetric solution, containing the LV parameter $\ell$, is obtained under a special potential $V(B^{\mu}B_{\mu}\pm b^{2})$ constructed from scalar combinations of the bumblebee vector field $B_{\mu}$ and the metric tensor $\mathrm{g}_{\mu\nu}$.  It is worth noticing that the choice of the potential has been done in such a way that it leads to a non-zero vacuum expectation value for $B_{\mu}$, including a spontaneous breaking of Lorentz symmetry \cite{CasanaPRD2018}. The non-vanishing vacuum expectation value for field $B_{\mu}$ depends on fulfilling the condition $B^{\mu}B_{\mu}\pm b^{2} = 0$. 
Indeed, this condition can be satisfied by considering $\langle B^{\mu} \rangle = b^{\mu}$, where the background field $b^{\mu}$ is a function of the spacetime coordinates so that $b^{\mu}b_{\mu} = \pm b^{2}=\text{constant}$. On the other hand,  the vector $b^{\mu}$ is responsible for the spontaneous breaking of Lorentz symmetry. 

Therefore, the corresponding vacuum solution induced by a spontaneous Lorentz symmetry violation is obtained as $\langle B^{\mu} \rangle$ remains frozen.  The bumblebee field $B_{\mu}$ is fixed to be $B_{\mu} = b_{\mu}$ which leads us to $V(B^{\mu}B_{\mu}\pm b^{2}) = 0$ and $V'(B^{\mu}B_{\mu}\pm b^{2}) = 0$
 \cite{CasanaPRD2018,KosteleckySamuel,Kostelecky,Bluhm,KosteleckPotting}. If $\ell (= \xi b^{2})$ goes to zero, the standard Schwarzschild space-time metric can be recovered, where $\xi$ is known as the real coupling constant that controls the non-minimal Bumblebee Gravity interaction.

Now starting from Ref. \cite{XuJCAP2018}, we intend to present the space-time metric of the Schwarzschild bumblebee BH surrounded by the DM described by the density profile. Therefore, The space-time metric for this scenario would be
\begin{equation}\label{MetricSchwlike11}
	\mathrm{d}s^{2}=-\mathcal{A}(r)\mathrm{d}t^{2}+\mathcal{B}(r)^{-1}\mathrm{d}r^{2}+C(r)\left(\mathrm{d}\theta^{2}+\text{sin}^{2}\theta\mathrm{d}\varphi^{2}\right), 
\end{equation}
where the metric coefficient functions $\mathcal{A}(r)$ and $\mathcal{B}(r)$ can be defined as
\begin{subequations}
	\begin{align}
		&\mathcal{A}(r) = A(r)+F_{1}(r),\\ 
		&\mathcal{B}(r) = \frac{B(r)}{\left(1+\ell\right)}+\frac{F_{2}(r)}{\left(1+\ell\right)}.
	\end{align}
\end{subequations}
Moreover, the modified Einstein equation, associated with the DM-BH system in  presence of a LV scenario, may be rearranged by redefining the energy-momentum tensor $T^{\nu}_{\,\,\,\mu}$. For such a combinatorial system, the modified energy-momentum tensor is derived by taking into account an additional part responsible for the energy-momentum tensor of DM that is $(T^{\mu}_{\,\,\,\nu})_{\mathrm{DM}}$.
Thus, the relevant modified Einstein equation can be written as 
\begin{equation}\label{ModEinsteinEq}
	R^{\mu}_{\,\,\,\nu} -\frac12 \delta^{\mu}_{\,\,\,\nu}R = \kappa^{2}\left((T^{\mu}_{\,\,\,\nu})_{\mathrm{DM}}+(T^{\mu}_{\,\,\,\nu})_{\mathrm{Schw}}\right).
\end{equation}
With regard to Eqs. \eqref{MetricSchwlike11} and \eqref{ModEinsteinEq}, we arrive at
\begin{subequations}
	\begin{align}
		&\left(\frac{B(r)}{\left(1+\ell\right)}+\frac{F_{2}(r)}{\left(1+\ell\right)}\right)\left(\frac{1}{r^{2}}+\frac{1}{r}\frac{B'(r)+F_{2}'(r)}{B(r)+F_{2}(r)}\right) = \frac{B(r)}{\left(1+\ell\right)}\left(\frac{1}{r^{2}}+\frac{1}{r}\frac{B'(r)}{B(r)}\right)\label{coupledEq1},\\ 
		&\left(\frac{B(r)}{\left(1+\ell\right)}+\frac{F_{2}(r)}{\left(1+\ell\right)}\right)\left(\frac{1}{r^{2}}+\frac{1}{r}\frac{A'(r)+F_{1}'(r)}{A(r)+F_{1}(r)}\right) = \frac{B(r)}{\left(1+\ell\right)}\left(\frac{1}{r^{2}}+\frac{1}{r}\frac{A'(r)}{A(r)}\right)\label{coupledEq2}.
	\end{align}
\end{subequations}
From Eqs. \eqref{coupledEq1} and \eqref{coupledEq2}, we get to the following expressions for $\mathcal{A}(r)$ and $\mathcal{B}(r)$
\begin{subequations}
	\begin{align}
		\mathcal{A}(r) &= \mathrm{e}^{\int \left(\frac{B(r)}{B(r)-\frac{2M_{\mathrm{BH}}}{r}}\left(\frac{1}{r}+\frac{A'(r)}{A(r)}\right)-\frac{1}{r}\right)\mathrm{d}r},\\
		\mathcal{B}(r) &= \frac{B(r)}{\left(1+\ell\right)} -\frac{2M_{\mathrm{BH}}}{\left(1+\ell\right)r}.
	\end{align}
\end{subequations}
Next, the space-time  metric of the Schwarzschild bumblebee BH surrounded by  DM is given  by
\begin{equation}\label{MetricSchwlike3}
	\begin{split}
		\mathrm{d}s^{2} &=-\mathrm{e}^{\int \left(\frac{B(r)}{B(r)-\frac{2M_{\mathrm{BH}}}{r}}\left(\frac{1}{r}+\frac{A'(r)}{A(r)}\right)-\frac{1}{r}\right)\mathrm{d}r}\mathrm{d}t^{2}+\left(1+\ell\right)\left(B(r)-\frac{2M_{\mathrm{BH}}}{r}\right)^{-1}\mathrm{d}r^{2}\\
		&+r^{2}\left(\mathrm{d}\theta^{2}+\text{sin}^{2}\theta\mathrm{d}\varphi^{2}\right), 
	\end{split}
\end{equation}
It is worth noticing that the assumption $A(r)=1=B(r)$ leads us to the conclusion that the space-time reduces to the pure Schwarzschild bumblebee BH space-time in  absence of DM  effects.
To establish of space-time metric of the Schwarzschild bumblebee BH surrounded by any DM, it is sufficient to incorporate the function $A(r)$ corresponding to each DM profile in Eq. \eqref{MetricSchwlike11}. Therefore, if we assume that $A(r)=B(r)$, the result is $F_{1}(r)=F_{2}(r)=-2M_{\mathrm{BH}}/r$. 
The metric tensor associated with the metric in
Eq. \eqref{MetricSchwlike3} can be written as
\begin{equation}\label{MetricSchwlike4}
	\begin{split}
		\mathrm{g}_{tt} &=-A(r)+\frac{2M_{\mathrm{BH}}}{r},\qquad \mathrm{g}_{rr} =\left(1+\ell\right)\left(A(r)-\frac{2M_{\mathrm{BH}}}{r}\right)^{-1},\\
		\mathrm{g}_{\theta\theta} &=r^{2},\qquad\qquad\qquad\qquad
		\mathrm{g}_{\varphi\varphi} =r^{2}\text{sin}^{2}\theta.
	\end{split}
\end{equation}


\section{ The Kerr bumblebee black hole in dark matter distribution
\label{sec3}
}

Applying the well-known method, called the Newman-Janis algorithm (NJA) \cite{NewmanJMP1965}, we can transform the static spherically symmetric space-time metric of the bumblebee BH immersed in DM into the rotational form.
Some authors by considering the complexification of coordinates have improved  the standard NJA method \cite{AzregPRD2014,AzregEPJC2014,ToshmatovPRD2014,ToshmatovEPJP2017}. Besides, some valuable studies have been developed in the context of  rotating DM-BH system by using this modified NJA method \cite{JusufiPRD2019,JusufiEPJC2020,JusufiMNRA2021,PantigJCAP2022, Arturo}.
To implement such a rotating DM-BH scenario,  we need to embed the spin parameter $a$ into the metric given by Eqs. \eqref{MetricSchwlike4}. To do so, relying on the modified NJA method, we convert the Boyer-Lindquist (BL) coordinates 
$\{t, r, \theta, \varphi\}$ to the Eddington-Finkelstein (EF) coordinates $\{u, r, \theta, \varphi\}$ by introducing the coordinate
$\mathrm{d}u = \mathrm{d}t - 1/\sqrt{\mathcal{A}(r)\mathcal{B}(r)}\mathrm{dr}$.
Thus, the metric in  Eq. \eqref{MetricSchwlike11} can be rewritten as 
\begin{equation}\label{MetricSchwlike2}
	\mathrm{d}s^{2}=-\mathcal{A}(r)\mathrm{d}u^{2}-2\sqrt{\frac{\mathcal{A}(r)}{\mathcal{B}(r)}}\mathrm{d}r\mathrm{d}u+C(r)\left(\mathrm{d}\theta^{2}+\text{sin}^{2}\theta\mathrm{d}\varphi^{2}\right), 
\end{equation}
In the null tetrad, the relevant space-time metric can be expressed by the linear combinations of four basis vectors $l^{\mu}$, $n^{\mu}$, $m^{\mu}$ and $\bar{m}^{\mu}$ and also the contravariant metric tensor is
\begin{equation}\label{nullTetrametrictensor1}
	\mathrm{g}^{\mu\nu} = -l^{\mu}n^{\nu}-l^{\nu}n^{\mu}+m^{\mu}\bar{m}^{\nu}+m^{\nu}\bar{m}^{\mu}, 
\end{equation}
where the four-basis vectors for the Schwarzschild bumblebee BH immersed in DM are
\begin{equation}\label{fourvec1}
	\begin{split}
		l^{\mu} &= \delta^{\mu}_{r}, \qquad\qquad\qquad\qquad\qquad\,\, n^{\mu} = \sqrt{\frac{\mathcal{B}}{\mathcal{A}}}\delta^{\mu}_{u}-\frac{\mathcal{A}}{2}\delta^{\mu}_{r},\\
		m^{\mu} &= \frac{1}{\sqrt{2C}}\left(\delta^{\mu}_{\theta}+\frac{i}{\text{sin}\theta}\delta^{\mu}_{\varphi}\right), \qquad 
		\bar{m}^{\mu} = \frac{1}{\sqrt{2C}}\left(\delta^{\mu}_{\theta}-\frac{i}{\text{sin}\theta}\delta^{\mu}_{\varphi}\right).
	\end{split}
\end{equation}
 The  null-tetrad  vectors $\vec{l}=l^{\mu}\partial_{\mu}$ and $\vec{n}=n^{\mu}\partial_{\mu}$
are real vectors, and $\vec{m}=m^{\mu}\partial_{\mu}$ and $\vec{\bar{m}}=\bar{m}^{\mu}\partial_{\mu}$ are complex vector as well. In other words, we are allowed to define $\vec{\bar{m}}$ as a complex conjugate of $\vec{m}$.
Furthermore, these vectors satisfy the following three conditions
\begin{subequations}
	\begin{align}
		&l^{\mu}l_{\mu} = n^{\mu}n_{\mu} = m^{\mu}m_{\mu} = \bar{m} ^{\mu} \bar{m}_{\mu} = 0, \label{vecCond1}\\
		&l^{\mu}m_{\mu} = l^{\mu}\bar{m}_{\mu} = n^{\mu}m_{\mu} = n^{\mu} \bar{m}_{\mu} = 0, \label{vecCond2}\\
		& l^{\mu}n_{\mu} =1, \quad  m^{\mu}\bar{m}_{\mu} =-1, \label{vecCond3}
	\end{align}
\end{subequations}
so that Eqs. \eqref{vecCond1}, \eqref{vecCond2} and \eqref{vecCond3} correspond to fulfill isotropy, orthogonality and normalization conditions, respectively. 
The non-null components of the contravariant metric tensor $\mathrm{g}^{\mu\nu}$ of the space-time metric for the Schwarzschild bumblebee BH, surrounded by DM in EF coordinates, are 
\begin{equation}
	\begin{split}
		& \mathrm{g}^{ur} = \mathrm{g}^{ru} = -\sqrt{\frac{\mathcal{B}}{\mathcal{A}}} = - \left(1+\ell\right)^{-1/2},\quad \mathrm{g}^{rr} = \mathcal{A}, \quad \mathrm{g}^{\theta\theta} = \frac{1}{C}, \quad \mathrm{g}^{\varphi\varphi} = \frac{1}{C \,\text{sin}^{2}\theta}.
	\end{split}
\end{equation}
Then, we need to perform complex coordinate transformations 
\begin{equation}
	u \rightarrow u-i a\, \text{cos}\theta,\qquad r \rightarrow r+i a \,\text{cos}\theta
\end{equation}
which means a rotation in Schwarzschild coordinates. This transformation procedure can be completed by taking $\delta^{\mu}_{\nu}$ as vectors as well as transforming them as
\begin{equation}
	\begin{split}
		&\delta^{\mu}_{u} \rightarrow \delta^{\mu}_{u},\qquad \qquad \qquad \qquad\qquad \delta^{\mu}_{r} \rightarrow \delta^{\mu}_{r},\\
		&\delta^{\mu}_{\theta} \rightarrow \delta^{\mu}_{\theta} +i a  \left(\delta^{\mu}_{u}-\delta^{\mu}_{r}\right) \text{sin}\theta, \qquad \delta^{\mu}_{\varphi} \rightarrow \delta^{\mu}_{\varphi}.
	\end{split}
\end{equation}
As a result of this procedure,  the metric coefficient  functions $\mathcal{A}(r)$, $\mathcal{B}(r)$ and $C(r)$, associated with each DM profile presented in this work,
are transformed into $\mathcal{\tilde{A}}(r,\theta,a)$, $\mathcal{\tilde{B}}(r,\theta,a)$ and $\Psi(r,\theta,a)$, respectively. 
Given $r \bar{r} = \tilde{C} = r^{2} +a^{2} \,\text{cos}^{2}\theta$, two of the new (transformed) functions, $\mathcal{\tilde{A}}(r,\theta,a)$ and $\mathcal{\tilde{B}}(r,\theta,a)$, can be rearranged by complexifying the radial coordinate $r$ as: 
\begin{itemize}
	\item for the DMS density profile
	\begin{equation}\label{DMSBHmetricCoeffFuncA1}
		\begin{split}
			\tilde{\mathcal{A}}_{\mathrm{DMS}}(r,\theta,a) &= \tilde{C}^{\,\frac{24\pi\rho_{\mathrm{R}}R_{\mathrm{SP}}^{\gamma_{\mathrm{SP}}}\left(\mathcal{K}R_{S}\right)^{3-\gamma_{\mathrm{SP}}}}{\gamma_{\mathrm{SP}}\left(\gamma_{\mathrm{SP}}-1\right)\left(\gamma_{\mathrm{SP}}-2\right)\left(\gamma_{\mathrm{SP}}-3\right)}} \,\,\mathrm{exp}\left(\frac{8\pi\rho_{\mathrm{R}}R_{\mathrm{SP}}^{\gamma_{\mathrm{SP}}}}{\left(\gamma_{\mathrm{SP}}-3\right)^{2}}\, \tilde{C}^{\,\frac{3-\gamma_{\mathrm{SP}}}{2}}\right.\\
			&\left.
			-\frac{24\pi\rho_{\mathrm{R}}R_{\mathrm{SP}}^{\gamma_{\mathrm{SP}}}\mathcal{K}R_{S}}{\left(\gamma_{\mathrm{SP}}-2\right)^{2}}\,\tilde{C}^{\,\frac{2-\gamma_{\mathrm{SP}}}{2}}+\frac{24\pi\rho_{\mathrm{R}}R_{\mathrm{SP}}^{\gamma_{\mathrm{SP}}}\left(\mathcal{K}R_{S}\right)^{2}}{\left(\gamma_{\mathrm{SP}}-1\right)^{2}}\,\tilde{C}^{\,\frac{1-\gamma_{\mathrm{SP}}}{2}}\right.\\
			&\left.
			-\frac{8\pi\rho_{\mathrm{R}}R_{\mathrm{SP}}^{\gamma_{\mathrm{SP}}}\left(\mathcal{K}R_{S}\right)^{3}}{\gamma_{\mathrm{SP}}^{2}}\,\tilde{C}^{\,-\frac{\gamma_{\mathrm{SP}}}{2}}
			\right)-\frac{2M_{\mathrm{BH}}\,r}{\tilde{C}},
		\end{split}
	\end{equation}	
	and 
	\begin{equation}\label{DMSBHmetricCoeffFuncB1}
		\tilde{\mathcal{B}}_{\mathrm{DMS}}(r,\theta,a) =  \frac{\tilde{\mathcal{A}}_{\mathrm{DMS}}(r,\theta,a)}{\left(1+\ell\right)},
	\end{equation}
	
	\item for the CDM density profile
	\begin{equation}\label{CDMBHmetricCoeffFuncA1}
		\begin{split}
			\tilde{\mathcal{A}}_{\mathrm{CDM}}(r,\theta,a) &= \left(1+\frac{\sqrt{\tilde{C}}}{R_{c}}\right)^{-\frac{8\pi \rho_{c} R_{c}^{3} r}{\tilde{C}}} -\frac{2M_{\mathrm{BH}}\,r}{\tilde{C}},
		\end{split}
	\end{equation}	
	and 
	\begin{equation}\label{CDMBHmetricCoeffFuncB1}
		\tilde{\mathcal{B}}_{\mathrm{CDM}}(r,\theta,a) =  \frac{\tilde{\mathcal{A}}_{\mathrm{CDM}}(r,\theta,a)}{\left(1+\ell\right)},
	\end{equation}

	\item for the BEC density profile
	\begin{equation}\label{TFBHmetricCoeffFuncA1}
		\begin{split}
			\tilde{\mathcal{A}}_{\mathrm{TF}}(r,\theta,a) &= \mathrm{e}^{-\frac{8\rho_{c} R_{c}^{3} r}{\pi^{2}\tilde{C}}\text{sin}\left(\frac{\pi \sqrt{\tilde{C}}}{R_{c}}\right)} -\frac{2M_{\mathrm{BH}}\,r}{\tilde{C}},
		\end{split}
	\end{equation}	
	and 
	\begin{equation}\label{TFBHmetricCoeffFuncB1}
		\tilde{\mathcal{B}}_{\mathrm{TF}}(r,\theta,a) =  \frac{\tilde{\mathcal{A}}_{\mathrm{TF}}(r,\theta,a)}{\left(1+\ell\right)}.
	\end{equation}
	
\end{itemize}
Moreover, the new form of  null tetrads $(l^{\mu}$, $n^{\mu}$, $m^{\mu})$, rearranged under such a transformation scheme, becomes 
\begin{equation}\label{Transnulltetrads}
	\begin{split}
		&\tilde{l}^{\mu} = \delta^{\mu}_{r}, \quad \tilde{n}^{\mu} = \left(1+\ell\right)^{-1/2}\delta^{\mu}_{u}-\frac{\mathcal{\tilde{A}}}{2}\delta^{\mu}_{r},\\
		&\tilde{m}^{\mu} = \frac{1}{\sqrt{2\Psi}}\left(\delta^{\mu}_{\theta} +i a  \left(\delta^{\mu}_{u}-\delta^{\mu}_{r}\right) \text{sin}\theta+\frac{i}{\text{sin}\theta}\delta^{\mu}_{\varphi}\right). 
	\end{split}
\end{equation}
Then, a composition of Eq. \eqref{nullTetrametrictensor1} and \eqref{Transnulltetrads} leads us to a new space-time metric stemming from the new covariant metric tensor $\tilde{\mathrm{g}}_{\mu\nu}$, so that
\begin{equation}\label{TransSTmetric1}
	\begin{split}
		\mathrm{d}\tilde{s}^{2} = \tilde{\mathrm{g}}_{uu}\mathrm{d}{u}^{2}+2\tilde{\mathrm{g}}_{ur}\mathrm{d}{u}\mathrm{d}{r} +2\tilde{\mathrm{g}}_{u\varphi}\mathrm{d}{u}\mathrm{d}{\varphi}+2\tilde{\mathrm{g}}_{r\varphi}\mathrm{d}{r}\mathrm{d}{\varphi}+\tilde{\mathrm{g}}_{\theta\theta}\mathrm{d}{\theta}^{2}+\tilde{\mathrm{g}}_{\varphi\varphi}\mathrm{d}{\varphi}^{2},
	\end{split}
\end{equation}
where 
\begin{equation}\label{CovMetricTansorCompon}
	\begin{split}
		& \tilde{\mathrm{g}}_{uu} = -\tilde{\mathcal{A}}, \qquad\qquad \tilde{\mathrm{g}}_{ur} = \tilde{\mathrm{g}}_{ru} = -\sqrt{1+\ell}, \\
		& \tilde{\mathrm{g}}_{u\varphi} =  \tilde{\mathrm{g}}_{\varphi u} = a \left(\tilde{\mathcal{A}}-\sqrt{1+\ell}\right)\text{sin}^{2}\theta,\qquad \tilde{\mathrm{g}}_{r \varphi} = \tilde{\mathrm{g}}_{\varphi r} = a \sqrt{1+\ell} \,\text{sin}^{2}\theta,\\
		& \tilde{\mathrm{g}}_{\theta\theta} = \Psi, \qquad\qquad \tilde{\mathrm{g}}_{\varphi\varphi} =\left(a^{2}\left(2\sqrt{1+\ell}-\tilde{\mathcal{A}}\right)\text{sin}^{2}\theta+\Psi\right)\text{sin}^{2}\theta.
	\end{split}
\end{equation}
Now, let us turn back the metric in Eq. \eqref{TransSTmetric1} from the EF coordinates to the BL coordinates by applying the following transformations
\begin{equation}\label{BLTrans}
	\mathrm{d}u = \mathrm{d}t-\frac{\left(1+\ell\right)^{-1/2}C+a^{2}}{\left(1+\ell\right)^{-1}\mathcal{A}\,C+a^{2}} \mathrm{d}r, \qquad \mathrm{d}\varphi = \mathrm{d}\varphi - \frac{a}{\left(1+\ell\right)^{-1}\mathcal{A}\,C+a^{2}} \mathrm{d}r.
\end{equation}
Based on considering the transformation given in Eq. \eqref{BLTrans}, the general form of the function $\mathcal{\tilde{A}}$ can be written in terms of the functions $\mathcal{A}$, $\Psi$ and $C$ as
\begin{equation}\label{NewFunctionA}
	\tilde{\mathcal{A}} = \frac{\left(\left(1+\ell\right)^{-1}\mathcal{A}\,C+a^{2}\text{cos}^{2}\theta\right)}{\left(\left(1+\ell\right)^{-1/2}C+a^{2}\text{cos}^{2}\theta\right)^{2}}\,\Psi,
\end{equation}
	Eventually, the general form of the rotating Schwarzschild bumblebee BH metric immersed in DM becomes
	\begin{equation}
		\begin{split}
			\mathrm{d}s^{2} &= -\frac{\left(\left(1+\ell\right)^{-1}\mathcal{A}\,C+a^{2}\text{cos}^{2}\theta\right)\Psi}{\left(\left(1+\ell\right)^{-1/2}C+a^{2}\text{cos}^{2}\theta\right)^{2}}\,\mathrm{d}{t}^{2}
			+\frac{\Psi}{\Delta}\,\mathrm{d}{r}^{2}\\
			&-2a \left(\frac{\left(1+\ell\right)^{-1/2}C-\left(1+\ell\right)^{-1}\mathcal{A}\,C}{\left(\left(1+\ell\right)^{-1/2}C+a^{2}\text{cos}^{2}\theta\right)^{2}}\right)\Psi\, \text{sin}^{2}\theta \,\mathrm{d}{\varphi}\mathrm{d}{t}+\Psi\,\mathrm{d}{\theta}^{2}\\
			&+\left(1+\frac{2\left(1+\ell\right)^{-1/2}C-\left(1+\ell\right)^{-1}\mathcal{A}\,C+a^{2}\text{cos}^{2}\theta}{\left(\left(1+\ell\right)^{-1/2}C+a^{2}\text{cos}^{2}\theta\right)^{2}}a^{2}\text{sin}^{2}\theta\right)\Psi\,\text{sin}^{2}\theta\,\mathrm{d}{\varphi}^{2},
		\end{split}
	\end{equation}
The general form of the rotating Schwarzschild bumblebee BH metric immersed in DM can be designed as the Kerr bumblebee BH metric 
\begin{subequations}\label{RotSTmet}
\begin{equation}\label{RotatingSTmetric1}
\begin{split}
\mathrm{d}s^{2} &= -\frac{\Psi}{\Sigma^{2}}\left(1-\frac{Q}{\Sigma^{2}}\right)\mathrm{d}{t}^{2}+\frac{\Psi}{\Delta}\,\mathrm{d}{r}^{2}-2a\, \frac{Q}{\Sigma^{4}}
\Psi\, \text{sin}^{2}\theta \,\mathrm{d}{t}\mathrm{d}{\varphi}
+\Psi\,\mathrm{d}{\theta}^{2}+\frac{\mathcal{G}\Psi\, \text{sin}^{2}\theta}{\Sigma^{4}}\mathrm{d}{\varphi}^{2},
\end{split}
\end{equation}
defining new notations
\begin{align}\label{QDeltaDM}
\mathrm{K}&	= \sqrt{\frac{\mathcal{B}}{\mathcal{A}}} =  \left(1+\ell\right)^{-1/2} C\\
\Sigma^{2} & = \mathrm{K}+a^{2}\text{cos}^{2}\theta,\\
Q &= \mathrm{K} - \left(1+\ell\right)^{-1} \mathcal{A} C,\\
\Delta &= \left(1+\ell\right)^{-1} \mathcal{A} C + a^{2},\\
\mathcal{G}& = \left(\mathrm{K}+a^2\right)^{2}-a^2 \Delta\, \text{sin}^{2}\theta.
\end{align}
\end{subequations}

Metric \eqref{RotSTmet} of a Kerr bumblebee BH has a general form, but its validity as a BH solution remains uncertain until the Einstein field equations are solved. To ensure its validity, the deformed BH metric \eqref{RotSTmet} must satisfy the conditions $G_{r\theta} = 0$ and $G_{\mu\nu} = 8\pi T_{\mu\nu}$ (see e.g. Refs. \cite{AzregPLB2014,AzregEPJC2014} for more detail), leading to the following equations
\begin{subequations}\label{ConformalFluidEq}
	\begin{align}
		&\left(\mathrm{K}+a^{2}y^{2}\right)^{2}\left(3\Psi_{,r}\Psi_{,y^{2}}-2\Psi\Psi_{,ry^{2}}\right) = 3a^{2} \mathrm{K}_{,r}\Psi^{2},\label{ConformalFluidEq1}\\
		&\Psi\left(\mathrm{K}_{,r}^{2}+\mathrm{K}\left(2-\mathrm{K}_{,rr}\right)-a^{2}y^{2}\left(2+\mathrm{K}_{,rr}\right)\right)+\left(\mathrm{K}+a^{2}y^{2}\right)\left(4y^{2}\Psi_{,y^{2}}-\mathrm{K}_{,r}\Psi_{,r}\right) =0,\label{ConformalFluidEq2}
	\end{align}
\end{subequations}
where $y\equiv \text{cos}\theta$. Note that, at present, $\Psi$ is an unknown function, and its determination is contingent upon satisfying the constraint $\mathcal{A}\ne\mathcal{B}$ or $\mathrm{K}\ne C=r^{2}$. This constraint arises due to the consideration of the DM-BH system in the presence of an LV scenario. Hence, a common solution, called conformal fluid, for complex partial differential Eqs. \eqref{ConformalFluidEq1} and \eqref{ConformalFluidEq2} can be expressed as follows:
\begin{equation}\label{CommSolutionPsi}
	\Psi=\mathrm{K}+a^{2}\text{cos}^{2}\theta,\qquad\qquad    \lim_{a \to 0} \Psi = \left(1+\ell\right)^{-1/2} r^2 \ne r^{2}.
\end{equation}
Taking into account the relevant considerations for this scenario, we can determine the  Kerr bumblebee BH metric as 
\begin{equation}\label{RotatingSTmetric2}
\begin{split}
\mathrm{d}s^{2} &= -\left(1-\frac{Q}{\Sigma^{2}}\right)\mathrm{d}{t}^{2}+\frac{\Sigma^{2}}{\Delta}\,\mathrm{d}{r}^{2}-2a\, \frac{Q}{\Sigma^{2}}\, \text{sin}^{2}\theta \,\mathrm{d}{t}\mathrm{d}{\varphi}
+\Sigma^{2}\,\mathrm{d}{\theta}^{2}+\frac{\mathcal{G}\, \text{sin}^{2}\theta}{\Sigma^{2}}\mathrm{d}{\varphi}^{2}.
\end{split}
\end{equation}
Let us now rearrange the new notations  $Q$ and $\Delta$
with regard to the relation  $\mathcal{M}_{\mathrm{DM-BH}}(r)=M_{\mathrm{DM}}+M_{\mathrm{BH}}= r(1-\mathcal{A})/2$, so that it is useful to rewrite $Q$ and $\Delta$ based on any DM profile discussed above \footnote{Here, in order to compress the notation,  instead of incorporating the mass corresponding to each DM profile shown in Eqs. \eqref{MassDMS}, \eqref{MassCDM} and \eqref{MassTF}, we simply set the corresponding symbols $\mathcal{M}_{\mathrm{DMS-BH}}(r)=M_{\mathrm{DMS}}+M_{\mathrm{BH}}$, $\mathcal{M}_{\mathrm{CDM-BH}}(r)=M_{\mathrm{CDM}}+M_{\mathrm{BH}}$ and $\mathcal{M}_{\mathrm{TF-BH}}(r)=M_{\mathrm{TF}}+M_{\mathrm{BH}}$ into the new symbols $\Delta$ and $Q$ assigned to each DM profile, respectively.}. 
Accordingly,  we have:
\begin{itemize}
	\item for the DMS density profile
	\begin{subequations}\label{QDeltaDMS}
		\begin{align}
			Q_{\mathrm{DMS}} &= \mathrm{K} - \left(1+\ell\right)^{-1} \left(r^{2}-2\mathcal{M}_{\mathrm{DMS-BH}}\, r\right),\\
			\Delta_{\mathrm{DMS}} &= \left(1+\ell\right)^{-1} \left(r^{2}-2\mathcal{M}_{\mathrm{DMS-BH}}\, r\right) + a^{2}\label{DeltaDMS},\\
			\mathcal{G}_{\mathrm{DMS}} & = \left(\mathrm{K}+a^2\right)^{2}-a^2 \Delta_{\mathrm{DMS}} \, \text{sin}^{2}\theta ,
		\end{align}
	\end{subequations}
	
	\item for the CDM density profile
	\begin{subequations}\label{QDeltaCDM}
		\begin{align}
			Q_{\mathrm{CDM}} &= \mathrm{K} - \left(1+\ell\right)^{-1} \left(r^{2}-2\mathcal{M}_{\mathrm{CDM-BH}}\, r\right),\\
			\Delta_{\mathrm{CDM}} &= \left(1+\ell\right)^{-1} \left(r^{2}-2\mathcal{M}_{\mathrm{CDM-BH}}\, r\right) + a^{2}\label{DeltaCDM},\\
			\mathcal{G}_{\mathrm{CDM}} & = \left(\mathrm{K}+a^2\right)^{2}-a^2 \Delta_{\mathrm{CDM}} \, \text{sin}^{2}\theta ,
		\end{align}
	\end{subequations}
	
	\item for the BEC density profile
	\begin{subequations}\label{QDeltaTF}
		\begin{align}
			Q_{\mathrm{TF}} &= \mathrm{K} - \left(1+\ell\right)^{-1} \left(r^{2}-2\mathcal{M}_{\mathrm{TF-BH}}\, r\right),\\
			\Delta_{\mathrm{TF}} &= \left(1+\ell\right)^{-1} \left(r^{2}-2\mathcal{M}_{\mathrm{TF-BH}}\, r\right) + a^{2}\label{DeltaTF},\\
			\mathcal{G}_{\mathrm{TF}} & = \left(\mathrm{K}+a^2\right)^{2}-a^2 \Delta_{\mathrm{TF}} \, \text{sin}^{2}\theta ,
		\end{align}
	\end{subequations}
	
\end{itemize}
This means that, by substituting each of the sets of Eqs. \eqref{QDeltaDMS}, \eqref{QDeltaCDM} and \eqref{QDeltaTF} into the  metric given in Eq. \eqref{RotatingSTmetric2}, the Kerr bumblebee BH metric describes the influence of the corresponding DM profiles and the spontaneous breaking of Lorentz symmetry  on the BH background. 
The space-time metric of the Kerr bumblebee BH surrounded by the DM distribution described by the DMS, CMD, and TF profiles is derived using Xu et al. method \cite{XuJCAP2018}. The metric given in Eq. \eqref{metricSchwlike1}, in the limit $\ell \rightarrow 0$, goes over the standard Schwarzschild BH metric.  In this limit, only the presence of DM particles in the vicinity of the BH can deform the Kerr BH metric describing the DM-BH interacting system.
In this case, we expect that as $\rho_{R}$ and $\rho_{c}$ go to zero, there is no DM distribution in the spike region and beyond. Hence, the LV spherically symmetric solution, that is the  Schwarzschild bumblebee BH solution generated by the bumblebee Gravity no longer is deformed  under the effect of  nearby DM. 
\section{Horizon and ergo-region of the Kerr bumblebee black hole in the dark matter spike
\label{sec4}
}

At this stage, we are interested in examining the variation of the horizon and the ergo-region of the Kerr bumblebee BH generated by the background with the spontaneous breaking of Lorentz symmetry surrounded by the DM distribution in the spike region.
Substituting Eqs. \eqref{QDeltaDMS}
into   metric   \eqref{RotatingSTmetric2} gives rise to a new metric where the influence of DMS distribution around  the Kerr bumblebee BH is evident. 
The impact of DM on the horizon and ergo-region depends on the distribution of mass in close proximity to the BH at the galactic center. As a result, we can disregard the DM  located far from the spike, specifically in regions where $r$ is significantly greater than $R_{\mathrm{SP}}$. Additionally, we have established that $\gamma_{\mathrm{SP}}$ falls within the range  $(0\div 3)$. It is worth noticing  that, at $\gamma_{\mathrm{SP}}=3$, there is a singularity. Typically, values of $\gamma_{\mathrm{SP}}$ greater than $2$ are not observed at the center of galaxies \cite{NampalliwarAJ2021}. 
It follows that the horizon and ergo-region depend on the values of the LV parameter $\ell$ and the parameters related to the presence of DM distributions around the BH. 

In this framework, let us take into account  observational data of the supermassive BH  at the center of M87 galaxy for the DMS profile.
As reported in \cite{XuJCAP2021,EventHorizon,JusufiPRD2019}, the values of DM halo parameters around the BH in M87 are $r_{0}=91.2\,\text{kpc}$ and $\rho_{0}=6.9\times10^{6}\,(M_{\odot}/\text{kpc}^{3})$. Given the BH M87 mass  $M_{\mathrm{BH}} = 6.5\times 10^{9}M_{\odot}$,  the corresponding Schwarzschild radius can be estimated as $R_{\mathrm{S}} \simeq 6.31 \times 10^{-7}\text{kpc}$. 

\begin{figure}[h]
	\centering 
	\includegraphics[width=.30\textwidth]{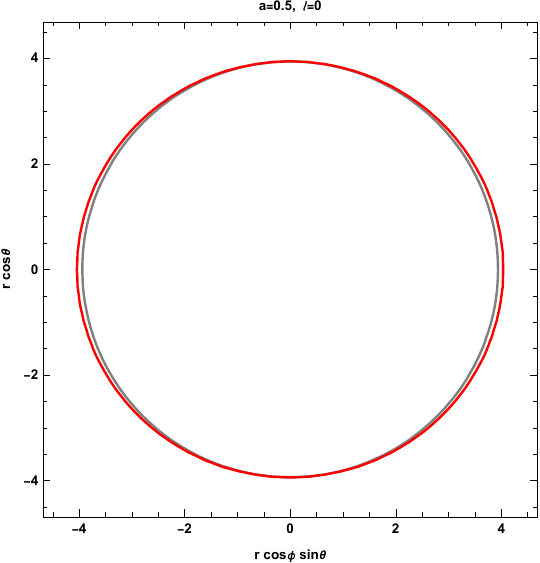}
	\hfill
	\includegraphics[width=.30\textwidth]{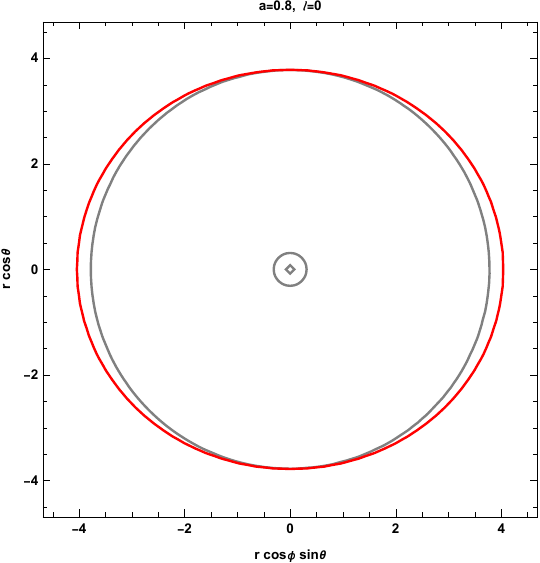}
	\hfill
	\includegraphics[width=.30\textwidth]{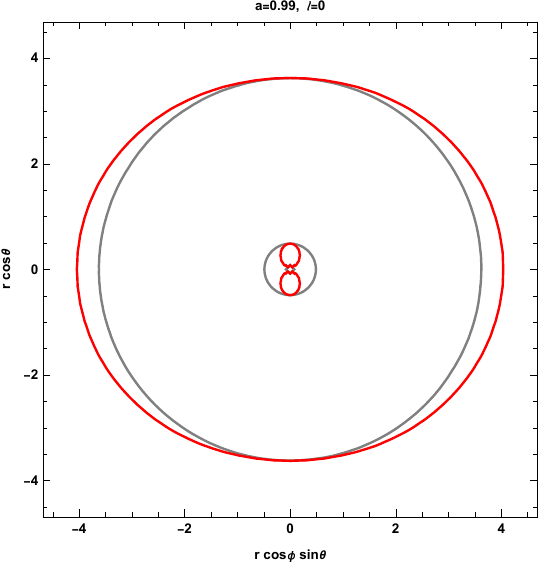}
	\hfill
	\includegraphics[width=.30\textwidth]{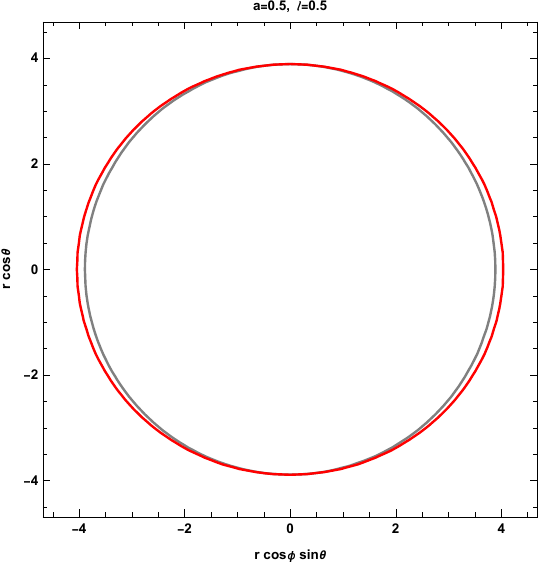}
	\hfill
	\includegraphics[width=.30\textwidth]{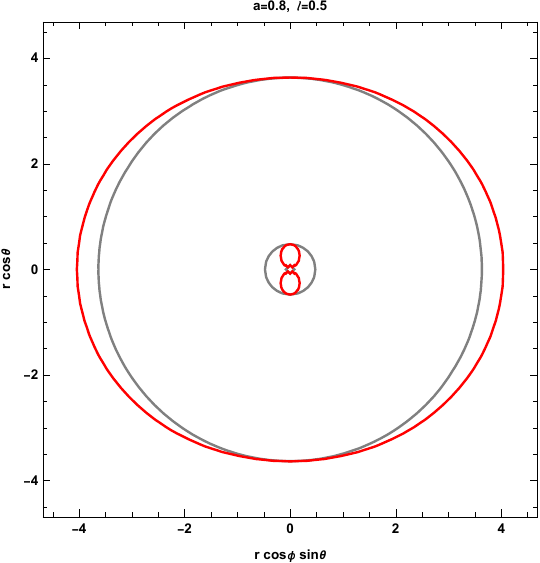}
	\hfill
	\includegraphics[width=.30\textwidth]{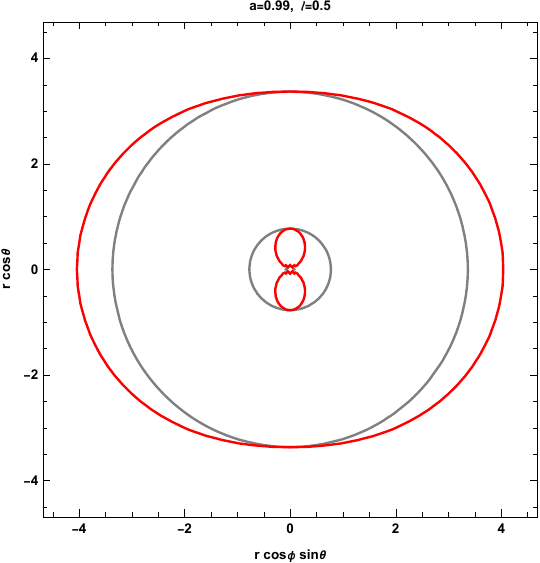}
	\hfill
	\includegraphics[width=.30\textwidth]{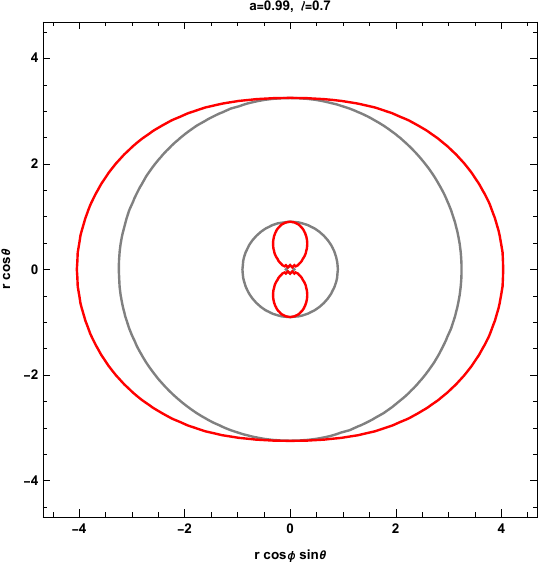}
	\hfill
	\includegraphics[width=.30\textwidth]{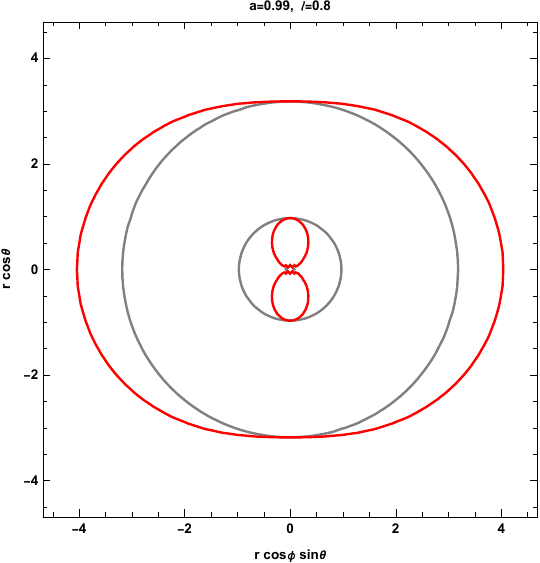}
	\hfill
	\includegraphics[width=.30\textwidth]{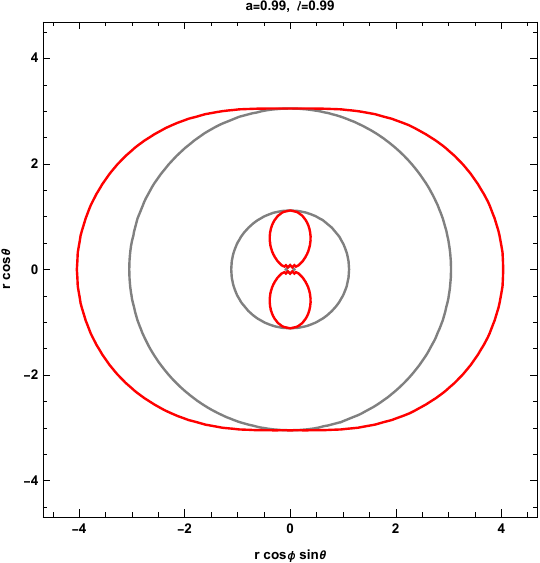}
	\caption{\label{fig:1} Visualizing the behavior of the ergo-region in the xz-plane of the Kerr bumblebee BH surrounded by DMS distribution with the red and grey lines assigned to the static limit surface and horizons, respectively. Here, in the first row, we assume $\ell=0$ versus three different values of the spin parameter, which is, $a=0.5$, $a=0.8$ and $a=0.99$; in the second row, we assume $\ell=0.5$ versus $a=0.5$, $a=0.8$ and $a=0.99$, and then in the third row, we assume  $a=0.99$ versus $\ell=0.7$, $\ell=0.8$ and $\ell=0.99$.
		}
\end{figure}

\begin{figure}[h]
	\centering 
	\includegraphics[width=.30\textwidth]{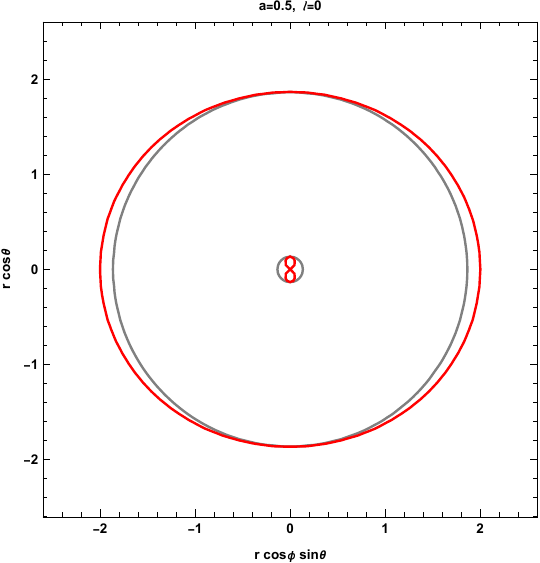}
	\hfill
	\includegraphics[width=.30\textwidth]{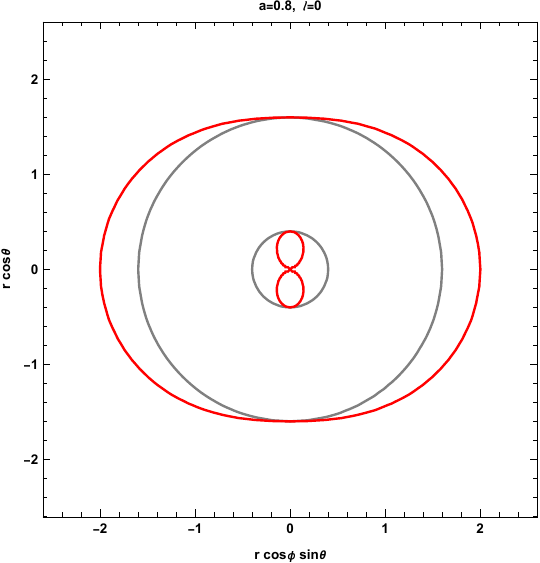}
	\hfill
	\includegraphics[width=.30\textwidth]{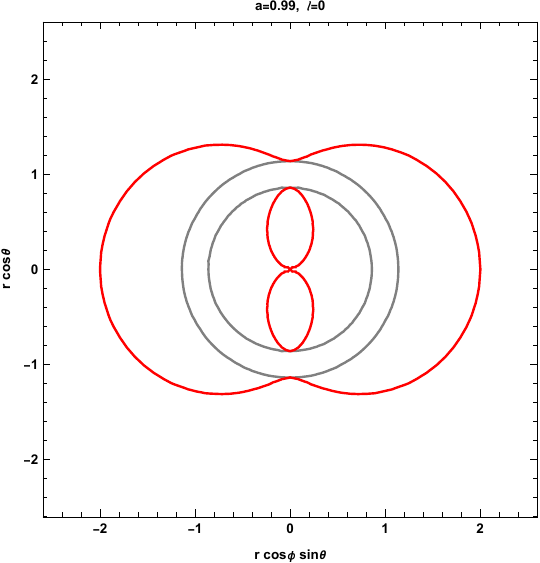}
	\hfill
	\includegraphics[width=.30\textwidth]{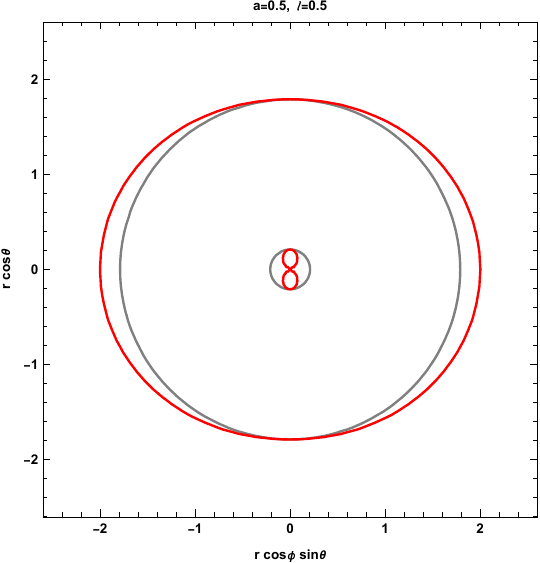}
	\hfill
	\includegraphics[width=.30\textwidth]{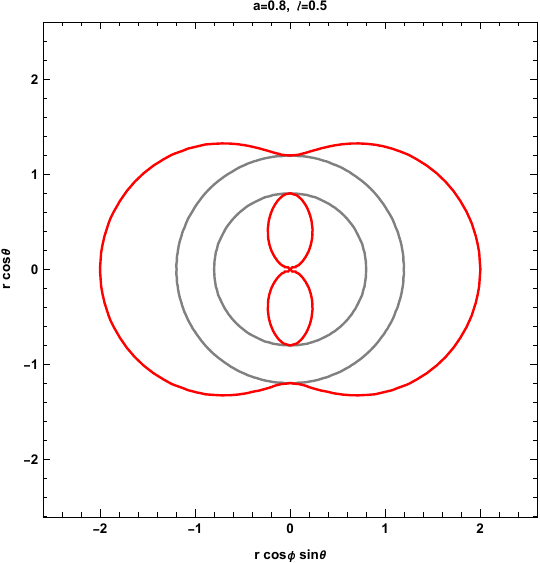}
	\hfill
	\includegraphics[width=.30\textwidth]{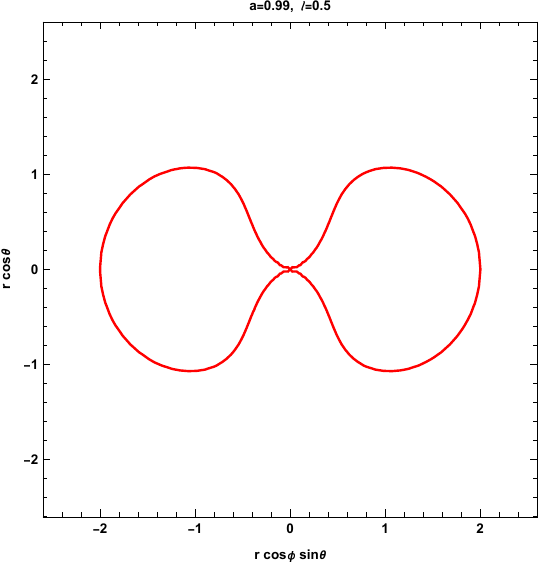}
	\hfill
	\includegraphics[width=.30\textwidth]{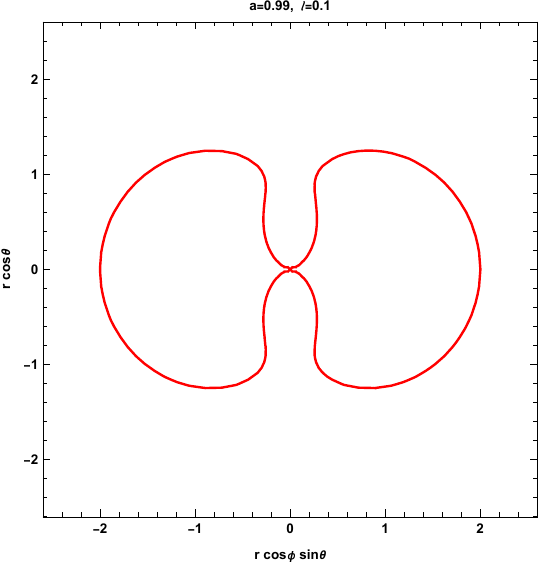}
	\hfill
	\includegraphics[width=.30\textwidth]{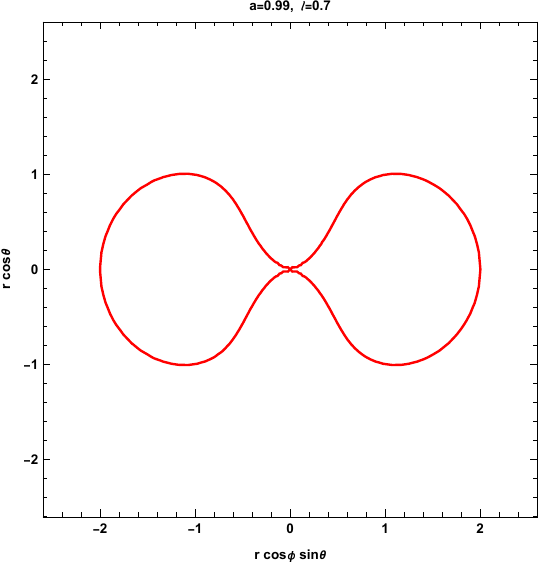}
	\hfill
	\includegraphics[width=.30\textwidth]{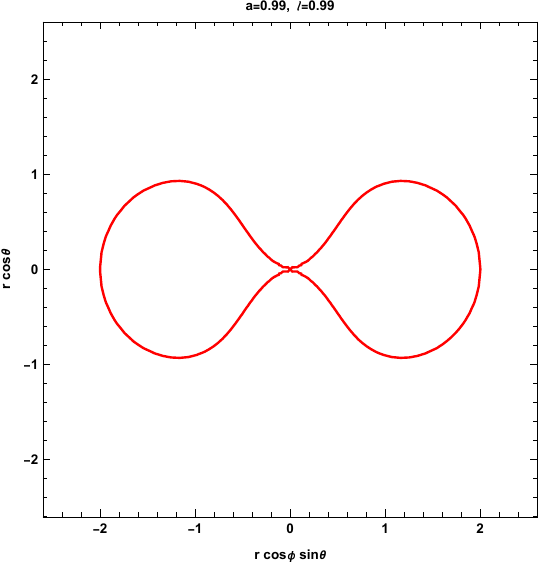}
	\caption{\label{fig:2} Visualizing the behavior of the ergoregion in the xz-plane of the Kerr bumblebee BH in the absence of DM distribution with the red and grey lines assigned to the static limit surface and horizons, respectively. Here, in the first row, we take $\ell=0$ versus three different values of the spin parameter, which is, $a=0.5$, $a=0.8$ and $a=0.99$; in the second row, we take $\ell=0.5$ versus $a=0.5$, $a=0.8$ and $a=0.99$, and then in the third row, we take $a=0.99$ versus $\ell=0.1$, $\ell=0.7$ and $\ell=0.99$.
	}
\end{figure}

It is widely known that the existence of a DM halo does not alter  BH horizons, such as the Kerr bumblebee BH, which possesses two horizons: the Cauchy horizon and the event horizon. The latter is referred to as the "surface of no return." With respect to Eqs. \eqref{RotatingSTmetric2} and \eqref{QDeltaDMS}, the identification of the horizon positions can be achieved through
\begin{subequations}\label{Horizon1}
	\begin{align}
		\left(1+\ell\right)^{-1} \left(r^{2}-2\mathcal{M}_{\mathrm{DMS-BH}}\, r\right) + a^{2}=0.
	\end{align}
\end{subequations}
An additional significant surface of a black hole is the  {\it static limit surface}. A defining characteristic of the static limit surface is that the character of particle geodesics undergoes a transformation upon crossing it, with a time-like geodesic becoming space-like and vice versa \cite{AmirEPJC2016}. The static limit surface meets the condition  $\mathrm{g}_{tt} = 0$, meaning that
\begin{equation}\label{limitsurface}
	\left(1+\ell\right)^{-1} \left(r^{2}-2\mathcal{M}_{\mathrm{DMS-BH}}\, r\right) +a^{2}\text{cos}^{2}\theta=0,
\end{equation}
However, due to Eq. \eqref{MassDMS}, a straightforward analytical solution for Eqs. \eqref{Horizon1} and \eqref{limitsurface} is not feasible. Therefore, through numerical plotting as demonstrated in Figs. \ref{fig:1} and \ref{fig:2}, the position of the horizons and the static limit surfaces is determined in the presence and absence of DM distribution, respectively.

Here, we depict the contours of the horizon surfaces in the xz-plane of the Kerr bumblebee BH in Figs. \ref{fig:1} and \ref{fig:2} to visually demonstrate the ergo-region as a function of varying $a$ and $\ell$ values. The ergo-region is defined by two boundaries: the event horizon and the outer static limit surface. An observer situated within the ergo-region cannot be static. The extent and shape of the ergo-region are influenced by the values of $a$ and $\ell$, and its area expands as these parameters increase. Besides, similar to the Cauchy horizon, there are observable deviations in the inner ergo-region.
In Figs. \ref{fig:1}, to indicate  behaviors of the  ergo-region, we consider the BH M87 surrounded by the DM distribution in the spike region with $ \mathcal{N}_{\gamma}=0.1$, $\gamma=1$, $\gamma_{\mathrm{SP}}=7/3$, $R_{\mathrm{S}} \simeq 6.31 \times 10^{-7}\text{kpc}$, $R_{\mathrm{SP}} \simeq 0.321 \text{kpc}$ and  $\rho_{\mathrm{R}}\simeq 0.301 (M_{\mathrm{BH}}/\text{kpc}^{3})$.
In Figs. \ref{fig:1} and \ref{fig:2}, we present the impact of the parameter $\ell$ on the ergo-region. The plots illustrate that the area of the ergo-region increases with both the rotation parameter $a$ and the LV parameter $\ell$. For BHs with faster rotation, the ergo-region is more sensitive to changes in the $\ell$ parameter. When $a$ approaches $1$, an increase in $\ell$ leads to a larger ergo-region. In fact, when $a$ approaches $1$, the impact of the $\ell$ parameter on the observable properties of the BH, such as horizons, the static limit surfaces, and ergo-regions, becomes more pronounced.
Besides, we can see that the presence of DMS around the BH causes both the event horizon and the outer static limit surface to expand. This suggests that the DMS has a significant effect on the properties of the BH gravitational field.
Here we note  that when the spin parameter $a$ and the LV parameter $\ell$  take on specific values, we observe that, in the absence of DMS, the Cauchy horizon and the inner static surface limit (which are opposite to the event horizon and the outer static limit surface, respectively) increase compared to when DMS is present.
It is important to mention that, whether or not there is a DMS, for a fixed spin parameter value, the Cauchy horizon expands while the event horizon contracts as the $\ell$ value increases.


\section{Shadow  of  the Kerr bumblebee black hole  in  the dark matter spike
\label{sec5}
}
In order to investigate the BH shadow phenomenon \cite{PerlickPR2022,HallaPRD2023,SolankiPRD2022,Cunha2015PRL,WeiJCAP2013,HiokiPRD2009,ShaikhPRD2019,AfrinMNRAS2021,HouJCAP2018,OvgunJCAP2018,TangJHEP2022}, it is crucial to attain a comprehensive understanding of the geodesic structure of a test particle within space-time metric in the BL coordinates given by Eqs. \eqref{RotatingSTmetric2} and \eqref{QDeltaDMS} with contravariant metric tensors
\begin{equation}\label{ModKerrContMetricTensor1}
\begin{split}
&\mathrm{g}^{tt} = -\frac{\mathcal{G} \Sigma^{2}}{\mathcal{G}\left(\Sigma^{2}-Q\right)+a^{2} Q^{2} \text{sin}^{2}\theta},\qquad \mathrm{g}^{t\varphi} =-\frac{2 a Q \Sigma^{2}}{\mathcal{G}\left(\Sigma^{2}-Q\right)+a^{2} Q^{2} \text{sin}^{2}\theta},\\
&\mathrm{g}^{rr} =\frac{\Delta}{\Sigma^{2}},\qquad\qquad\quad \mathrm{g}^{\theta\theta} = \frac{1}{\Sigma^{2}},\qquad \mathrm{g}^{\varphi\varphi} = \frac{\Sigma^{2}\left(\Sigma^{2}-Q\right)\text{csc}^{2}\theta}{\mathcal{G}\left(\Sigma^{2}-Q\right)+a^{2} Q^{2} \text{sin}^{2}\theta}.
\end{split}
\end{equation}
To achieve this objective, a Lagrangian formulation can be adopted.  After, one can use  the Hamilton-Jacobi equation and the Carter constant separable method \cite{CarterPR1968,Chandrasekhar}. The motion of a particle can be described by the Lagrangian 
\begin{equation}
\mathcal{L} = \frac12 \mathrm{g}_{\mu\nu} \dot{x}^{\mu}\dot{x}^{\nu},
\end{equation}
where the four-velocity of the particle is  $u^{\mu} = \dot{x}^{\mu} = x^{\mu}_{,\lambda}$ and $\lambda$ is the affine parameter  along the geodesic.
The conservation of conjugate momenta $p_{t}$ and $p_{\varphi}$ is a result of the symmetry of the BH, whereby metric-independent variables $t$ and $\varphi$ are involved. Additionally, it is worth noting that metric \eqref{RotatingSTmetric2} incorporates two Killing vectors, $\xi^{\mu} = (\frac{\partial}{\partial t})^{\mu}$ and $\chi^{\mu} = (\frac{\partial}{\partial \varphi})^{\mu}$, that correspond to two conserved quantities, namely the energy $E$ and the angular momentum $L$. This implies that energy $E$ and angular momentum $L$ can be computed as follows:
\begin{equation}\label{geoEq12}
\begin{split}
&E = - \mathrm{g}_{\mu\nu} \xi^{\mu} x^{\mu}_{,\lambda}=- p_{t} = - \frac{\partial \mathcal{L}}{\partial \dot{t}} = - \mathrm{g}_{tt} \dot{t} - \mathrm{g}_{t \varphi} \dot{\varphi},\\
&L =  \mathrm{g}_{\mu\nu} \chi^{\mu} x^{\mu}_{,\lambda}= p_{\varphi} =  \frac{\partial \mathcal{L}}{\partial \dot{\varphi}} =  \mathrm{g}_{\varphi t} \dot{t} + \mathrm{g}_{\varphi \varphi} \dot{\varphi},
\end{split}
\end{equation}
and lead us to the following expressions

\begin{subequations}
\begin{align}
& \Sigma^{2} \dot{t} =-a\left(a E\, \text{sin}^{2}\theta-L\right)+\frac{\mathcal{U}\left(E\,\mathcal{U}-a L\right)}{\Delta} , \label{geoEq1}\\
&\Sigma^{2} \dot{\varphi} =  -\left(a E -\frac{L}{\text{sin}^{2}\theta}\right) +\frac{a E\, \mathcal{U}-a^{2} L}{\Delta} ,\label{geoEq2}
\end{align}
\end{subequations}
where $\mathcal{U} \equiv (\left(1+\ell\right)^{-\frac12}r^{2}+a^2)$. By taking the limit as $\ell$ goes to $0$, the function $\mathcal{U}$ approaches to $(a^2 + r^2)$, which is the expression for the standard Kerr BH \cite{Kerr}.
In order to fully describe the dynamics of a photon in the vicinity of a Kerr BH under Bumblebee Gravity, subject to a distribution of DM, we must derive two additional null geodesic equations using the Hamilton-Jacobi method, which can be expressed as \cite{CarterPR1968,JohannsenPRD2013}
\begin{equation}\label{HJequation}
\frac12  \mathrm{g}^{\mu\nu} \frac{\partial\mathcal{S}}{\partial x^{\mu}} \frac{\partial\mathcal{S}}{\partial x^{\nu}} =- \frac{\partial\mathcal{S}}{\partial \lambda}.
\end{equation}
To obtain the separable solution of Eq. \eqref{HJequation}, we need to express the action $\mathcal{S}$ in the following form \cite{WeiJCAP2013,HiokiPRD2009,ShaikhPRD2019,AfrinMNRAS2021,HouJCAP2018,TangJHEP2022}:
\begin{equation}\label{action}
\mathcal{S} = \frac{1}{2} m^{2} \lambda -Et+L\varphi+\mathcal{S}_{r}(r)+\mathcal{S}_{\theta}(\theta), 
\end{equation}
with functions of the coordinates $r$ and $\theta$, i.e., $\mathcal{S}_{r}$ and $\mathcal{S}_{\theta}$, and the mass of the test particle $m$. However, for a photon, $m$ equals to zero. Substituting Eq. \eqref{action} into Eq. \eqref{HJequation} leads to derive the following geodesic equations 
\begin{subequations}
\begin{align}
\Delta \left(\frac{\partial\mathcal{S}_{r}}{\partial r}\right)^{2} &=\frac{\mathcal{R}(r)}{\Delta},\label{geoEq3}\\
\left( \frac{\partial\mathcal{S}_{\theta}}{\partial \theta}\right)^{2} &= \Theta(\theta),\label{geoEq4}
\end{align}
\end{subequations}
where 
\begin{subequations}
\begin{align}
&\mathcal{R}(r) =\left(E \,\mathcal{U}-a L\right)^{2}-\Delta\left(\mathcal{Q}+\left(a E-L\right)^{2}\right),\label{FunctionR}\\
&\Theta(\theta) = \mathcal{Q}+\left(a^{2}E^{2}-\frac{L^{2}}{\text{sin}^{2}\theta}\right)\text{cos}^{2}\theta .\label{Functiontheta}
	\end{align}
\end{subequations}
The Carter constant $\mathcal{Q}$ is defined as $\mathcal{Q} \equiv \tilde{k}-(a E-L)^{2}$, where $\tilde{k}$ is another constant of motion \cite{JusufiPRD2019,JusufiEPJC2020,JusufiMNRA2021,PantigJCAP2022,AzregPRD2014}. 
The behavior of photons near BHs is a topic of great complexity, which is extensively discussed  \cite{HiokiPRD2008,JohannsenPRD2011}. Notably, the photon motion is distinctly different from that outside a certain radius near the event horizon of the BH. Within this critical radius, there are no stable photon orbits, but outside  it, a stable orbit exists. Due to the strong gravitational field near the BH, it is expected that photons emitted in its vicinity will either fall into the BH or scatter away from it \cite{TangJHEP2022}.

A method to examine the presence of unstable circular orbits around the BH is to recast the radial geodesic equation in terms of the effective potential $V_{\text{eff}}\equiv \mathcal{R}(r)/E^{2}$ associated with the photon radial motion as
\begin{equation}\label{geoEq33}
\Delta^{2} \left(\frac{\partial\mathcal{S}_{r}}{\partial r}\right)^{2} +V_{\text{eff}}=0.
\end{equation}
In this way, the dimensionless parameters $\xi=L/E$ and $\eta = \mathcal{Q}/E^2$ can be defined to study the characteristics of unstable circular photon orbits in general rotating space-time \cite{NampalliwarAJ2021}. 
Our assumptions for this analysis include that both the observer and photons are positioned at infinity, and that the photons approach the equatorial plane with $\theta=\pi/2$.
Using these two parameters, $\xi$ and $\eta$, we can formulate the effective potential as
\begin{equation}
V_{\text{eff}} = \left(\mathcal{U}-a \xi\right)^{2}-\Delta\left(\eta+\left(a-\xi\right)^{2}\right),
\end{equation}
 It can be shown that these orbits must satisfy the conditions $V_{\text{eff}}(r)|_{r=r_{\mathrm{ph}}}=0$, $V_{\text{eff}}'(r)|_{r=r_{\mathrm{ph}}}=0$, and $V_{\text{eff}}''(r)|_{r=r_{\mathrm{ph}}}\le 0$ at the radius $r=r_{\mathrm{ph}}$ of the unstable orbit \cite{JohannsenAPJ2013}.
The conditions stated above allow us to demonstrate that the impact parameters $\xi$ and $\eta$ have critical values for unstable orbits, which can be expressed as 

\begin{subequations}\label{impactParameters}
\begin{align}
\xi & = \left. \frac{\mathcal{U}\Delta_{,r}-2\Delta\,\mathcal{U}_{,r}}{a \Delta_{,r}}\right|_{r\rightarrow r_{\mathrm{ph}}},\label{impactParameterXi}\\
\eta & =\left. \frac{4 a^{2} \,\mathcal{U}^{2}_{,r}\Delta-\left(\left(\,\mathcal{U}-a^{2}\right)\Delta_{,r}-2\,\mathcal{U}_{,r} \Delta\right)^{2}}{a^{2}\Delta^{2}_{,r}}\right|_{r\rightarrow r_{\mathrm{ph}}}.\label{impactParameterEta}
\end{align}
\end{subequations}
Utilizing the defined functions $\mathcal{U}$ and $\Delta$, we can now derive an expression for $\xi^2+\eta$ in terms of the mass of the DM-BH interacting system. Specifically, we obtain the following expression:
\begin{equation}\label{PlusImpactParameters}
\begin{split}
\xi^2+\eta = -a^2 +2\left(a^{2}+\frac{r^2}{\sqrt{1+\ell}}\right)+\frac{4r^2 \Delta (1+\ell)}{\left(\mathcal{M}+r\left(\mathcal{M}_{,r}-1\right)\right)^{2}}+\frac{4 r \Delta \sqrt{1+\ell}}{\left(\mathcal{M}+r\left(\mathcal{M}_{,r}-1\right)\right)}.
\end{split}
\end{equation}
It is worth noticing that the functions $\Delta$ and $\mathcal{M}$ are specific to each DM-BH interacting system in the context of Bumblebee Gravity, such that Eq. \eqref{PlusImpactParameters} can be reproduced using $(\Delta_{\text{DMS}}, \mathcal{M}_{\mathrm{DMS-BH}})$, $(\Delta_{\text{CDM}}, \mathcal{M}_{\mathrm{CDM-BH}})$, and $(\Delta_{\text{TF}}, \mathcal{M}_{\mathrm{TF-BH}})$ for each system. However, in this work, we shift our attention towards the Kerr bumblebee BH metric within a DMS.

Having obtained this information, we can employ the impact parameters $\xi$ and $\eta$ represented by Eqs. \eqref{impactParameterXi} and \eqref{impactParameterEta} to examine the shape of our BH shadow surrounded by photons that have escaped from unstable orbits.

After determining the geodesic of the photon, we can investigate the motion of the photon as observed by an observer located at position $(\tilde{r}_{0}, \theta_{0})$ relative to a BH. Here, $\tilde{r}_{0}$ denotes the distance between the observer and the BH, while $\theta_{0}$ represents the inclination angle defined as the angle between the rotation axis of the BH and the observer line of sight. As such, in the presence of a possible scenario of  Lorentz violation, we can examine the effect of a DMS on the BH shadow images.
To this aim, we need to apply a two-dimensional coordinate system on the observer plane called the celestial coordinate $\{\alpha, \beta\}$.
Indeed, the observer has the option to select a Cartesian coordinate system with the BH as its center. In this system, the spherical photon orbits can be projected onto the celestial plane to create a closed curve that is parameterized by the celestial coordinates $\{\alpha, \beta\}$. This curve serves to delineate the trajectory of the photons as they appear to the observer from their position relative to the BH \cite{KumaraJCAP2020}. In order to understand the geometry of the BH shadow, we need to map the photons emitted from its vicinity to a celestial coordinate system, where they correspond to individual coordinates $\{\alpha, \beta\}$. By tracing back the geodesic motion of each photon by its respective $\{\alpha, \beta\}$, we can determine its trajectory as seen by an observer located at some position relative to the BH. To create the BH shadow image, we must project these trajectories onto the image plane, which requires us to map the celestial coordinates to the BL coordinates of the BH. Thus, the relationship between the celestial coordinates and the BL coordinates, that is
\begin{subequations}\label{CelestialCoord1}
\begin{align}
\alpha = -\tilde{r}_{0} \frac{p^{(\varphi)}}{p^{(t)}}, \qquad \beta = -\tilde{r}_{0} \frac{p^{(\theta)}}{p^{(t)}},
\end{align}
\end{subequations} 
 allows us to fully understand the geometry of the BH shadow. Here, the components of the photon four-momentum associated with a locally non-rotating reference frame are denoted by $p^{(\mu)}$. 
 To find the tetrad components of $p^{(\mu)}$, we require an inertial reference frame of an observer located at a significant distance from the BH, or any local observer. The basis vectors $e_{(\nu)}$ of this reference frame are defined as $e_{(\nu)} = e^{\mu}_{(\nu)} e_{\mu}$, with the coordinate basis $(e_{t},e_{r},e_{\theta},e_{\varphi})$ of the metric and $e^{\mu}_{(\nu)} = (\zeta_{c},\gamma_{c}, \frac{1}{\sqrt{\mathrm{g}_{rr}}},\frac{1}{\sqrt{\mathrm{g}_{\theta\theta}}},\frac{1}{\sqrt{\mathrm{g}_{\varphi\varphi}}})$ \cite{JohannsenPRD2013,JohannsenAPJ2013}. Let us now expand the following components of $p^{(\mu)}$ in terms of the observer bases as 
\begin{subequations}\label{TetCompPhotonMom}
\begin{align}
& p^{(t)} =  \zeta_{c} E - \gamma_{c} L, \qquad\qquad\,\,\qquad\qquad p^{(\varphi)} = \frac{L}{\sqrt{\mathrm{g}_{\varphi\varphi}}},\\
& p^{(\theta)} = \frac{p_{\theta}}{\sqrt{\mathrm{g}_{\theta\theta}}} = \pm\frac{\sqrt{\Theta(\theta)}}{\sqrt{\mathrm{g}_{\theta\theta}}},\quad\quad\quad\qquad p^{(r)} = \frac{p_{r}}{\sqrt{\mathrm{g}_{rr}}},
\end{align}
\end{subequations} 
where $\zeta_{c}$  and $\gamma_{c}$ are real  constants and also $\Theta(\theta)\equiv \eta + a \text{cos}^{2}\theta - \xi^{2}\text{cot}^{2}\theta$.
As previously mentioned, the conserved quantities $E\equiv-p_{t}$ and $L\equiv p_{\varphi}$ are associated with the Killing vectors. 
To obtain the constants $\zeta_{c}$ and $\gamma_{c}$, one can utilize the orthonormal property of the system and the relation $e^{\mu}_{(\alpha)}e^{\nu}_{(\beta)}\mathrm{g}_{\mu\nu} = \eta_{\alpha\beta}$ with $\eta_{\alpha\beta}=\text{diag}(-+++)$. This yields:
\begin{subequations}
\begin{align}
\zeta_{c} = \sqrt{\frac{\mathrm{g}_{\varphi\varphi}}{\mathrm{g}^{2}_{t\varphi}-\mathrm{g}_{tt}\mathrm{g}_{\varphi\varphi}}},\qquad\qquad
\gamma_{c} = \frac{\mathrm{g}_{t\varphi}}{\mathrm{g}_{\varphi\varphi}}\sqrt{\frac{\mathrm{g}_{\varphi\varphi}}{\mathrm{g}^{2}_{t\varphi}-\mathrm{g}_{tt}\mathrm{g}_{\varphi\varphi}}},
\end{align}
\end{subequations}
Additionally, we can rewrite the celestial coordinates in terms of $\xi$ and $\eta$. 
The resulting expressions are as follows:
\begin{subequations}\label{CelestialCoord2}
\begin{align}
&\alpha = \left.-\tilde{r}_{0} \frac{\xi}{\sqrt{\mathrm{g}_{\varphi\varphi}}\zeta_{c}\left(1+\frac{\mathrm{g}_{t\varphi}}{\mathrm{g}_{\varphi\varphi}}\xi\right)}\right|_{(r\rightarrow \tilde{r}_{0}, \, \theta\rightarrow\theta_{0})},\\
&\beta = \left.\pm \tilde{r}_{0} \frac{\sqrt{ \eta + a \text{cos}^{2}\theta - \xi^{2}\text{cot}^{2}\theta}}{\sqrt{\mathrm{g}_{\theta\theta}}\zeta_{c}\left(1+\frac{\mathrm{g}_{t\varphi}}{\mathrm{g}_{\varphi\varphi}}\xi\right)}\right|_{(r\rightarrow \tilde{r}_{0}, \, \theta\rightarrow\theta_{0})},
\end{align}
\end{subequations} 
Moreover, in the case where the observer is located at the equatorial plane
with an inclination angle of $\theta_{0} = \frac{\pi}{2}$, and positioned at a considerably large, yet finite distance, of approximately $\tilde{r}_{0} \sim 16.8 \,\text{Mpc}$, expressions  can be simplified as follows \cite{NampalliwarAJ2021}:
\begin{equation}\label{CelestialCoord4}
	\alpha  = -\sqrt{A(\tilde{r}_{0})}\,\xi, \qquad\qquad \beta=\pm \sqrt{A(\tilde{r}_{0})}\,\sqrt{\eta}.
\end{equation}
We can observe, from the expression given in Eq. \eqref{DMSmetricCoeffFuncA1}, that $A(r)$ plays a crucial role in our analysis. It is important to note that due to the presence of the DMS, our solution is non-asymptotically flat. This contrasts with the asymptotically flat case, which occurs when the DMS is absent and $A(r)$ approaches $1$. For the observer located at $(\tilde{r}_{0}, \theta_{0})$ (i.e. the observer is far but not at infinity), we have neglected the effect of rotation while determining the values of $\xi$ and $\eta$ using Eq. \eqref{impactParameters} \cite{JusufiEPJC2020}.
Using the celestial coordinates $\{\alpha,\beta\}$, 
 we provide  an expression to visualize the shapes of  BH shadow, given by:
\begin{equation}
 \alpha^{2}+\beta^{2} = A(\tilde{r}_{0}) (\xi^{2}+\eta).
\end{equation}
We consider a collection of photons with all possible values, $\xi$ and $\eta$. It is evident that the unstable circular photon orbits create the border of a shadow, as photons inside these orbits are trapped and unable to escape to infinity. The $\xi$ and $\eta$ values of the unstable photon orbits are determined by the radial variable $r_{\mathrm{ph}}$, as specified in Eq. \eqref{impactParameterEta}. Thus, the boundary of the shadow, in the observer sky, is defined by parametric functions expressed in terms of the parameter $r_{\mathrm{ph}}$. Specifically, the functions $\alpha = \alpha(r_{\mathrm{ph}})$ and $\beta = \beta(r_{\mathrm{ph}})$ determine the location of the shadow boundary in the observer sky. The allowed range of $r_{\mathrm{ph}}$  is restricted by the condition $\mathcal{Q}\ge0$. 
Besides, when a BH is viewed from a specific angle $\theta_{0}$, its maximum distortion occurs at its highest possible angular momentum. Moreover, the largest distortion for a BH mass and angular momentum is observed from the equatorial plane at $\theta_{0} = \frac{\pi}{2}$, elongating the vertical $\beta$-direction while squeezing the horizontal $\alpha$-direction, while the shadows remain convex. There are two real solutions $\beta(\theta = \theta_{0})\equiv\eta(r_{\mathrm{ph}}^{\pm})=0$ that satisfy the condition $r_{\mathrm{ph}}^{+}\ge r_{\mathrm{ph}}^{-}$ for $ \beta= 0$. Thus, based on the definition provided in Refs. \cite{,NampalliwarAJ2021,FengEPJC2020}, the size of the shadow can be defined by the left-most and right-most coordinates, $\alpha(r_{\mathrm{ph}}^{-})$ and $\alpha(r_{\mathrm{ph}}^{+})$, and given by 
\begin{equation}
R_{\mathrm{sh}} = \frac12 \left(\alpha(r_{\mathrm{ph}}^{+})-\alpha(r_{\mathrm{ph}}^{-})\right).
\end{equation}
The trajectory of photons traveling from a BH to an observer located at $\tilde{r}_{0}$ is influenced by the geometry of the BH surroundings. Specifically, the radius of circular null geodesics, denoted as $r_{\mathrm{ph}}^{\pm}$, varies depending on whether the BH is spinning or static. In the case of a static metric and a BH with zero spin, $r_{\mathrm{ph}}^{\pm} = 3 M_{\mathrm{BH}}$ is obtained, which corresponds to the radius of the shadow $R_{\mathrm{sh}}=\sqrt{A(\tilde{r}_{0})}\,3\sqrt{3} M_{\mathrm{BH}}$.

\begin{figure}[h]
	\centering 
	\includegraphics[width=.45\textwidth]{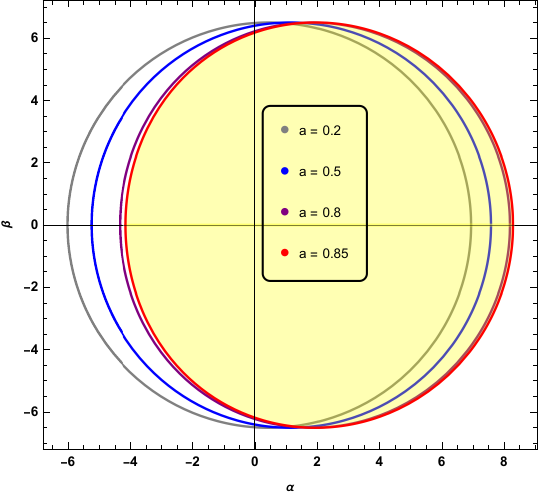}
	\hfill
	\includegraphics[width=.45\textwidth]{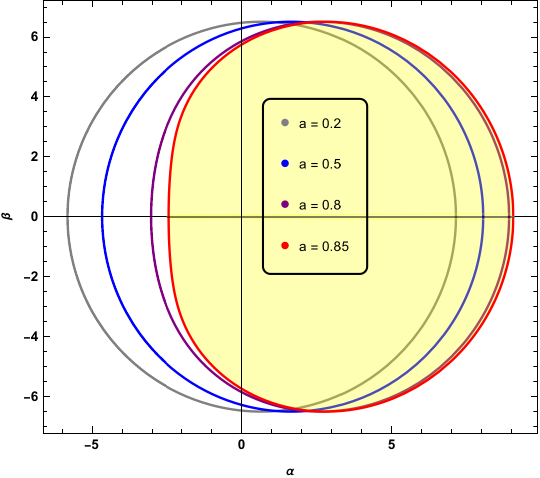}
	\hfill
	\includegraphics[width=.45\textwidth]{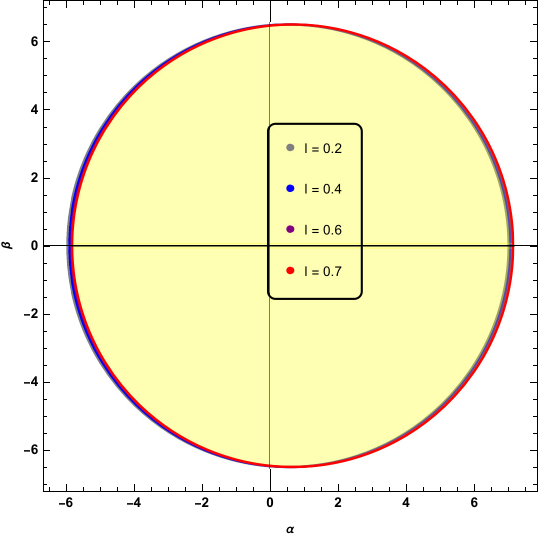}
	\hfill
	\includegraphics[width=.45\textwidth]{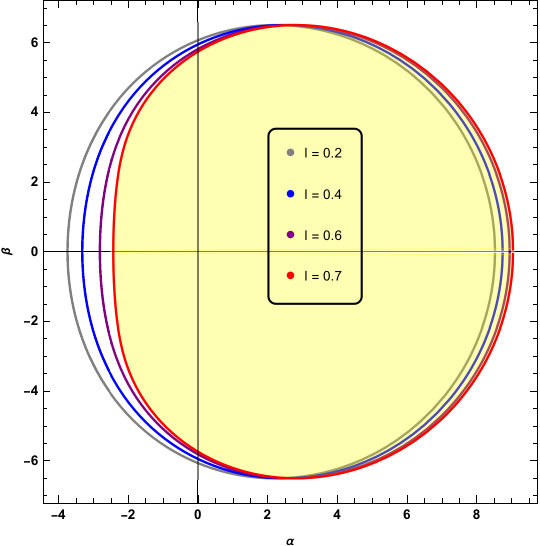}
	\caption{\label{fig:3}
Variation in the shadow shape of the bumblebee BH M87 surrounded by DM distribution using $ \mathcal{N}_{\gamma}=0.1$, $\gamma=1$, $\gamma_{\mathrm{SP}}=7/3$, 
$r_{0}\simeq1.71\,\text{kpc}$, $\rho_{0}\simeq 6.9\times10^{6}\,(M_{\odot}/\text{kpc}^{3})$,  $M_{\mathrm{BH}} \simeq 6.5\times 10^{9}M_{\odot}$
$R_{\mathrm{S}} \simeq 6.31 \times 10^{-7}\,\text{kpc}$, $R_{\mathrm{SP}} \simeq 2.347\, \text{kpc}$,  $\rho_{\mathrm{R}}\simeq 0.0008\, (M_{\mathrm{BH}}/\text{kpc}^{3})$ and $\tilde{r}_{0} \simeq 16.8 \,\text{Mpc}$
for different values of $a$ and $\ell$. We take $M_\mathrm{BH} = 1$  in units of the BH M87 mass.	
Here we have set $\text{kpc}=1$.  
The upper left panel corresponds to $\ell=0$ and $a=0.2, 0.5, 0.8, 0.85$, the upper right panel corresponds to $\ell=0.7$ and $a=0.2, 0.5, 0.8, 0.85$, the lower left panel corresponds to $a=0.2$ and $\ell=0.2, 0.4, 0.6, 0.7$, and the lower right panel corresponds to $a=0.85$ and $\ell=0.2, 0.4, 0.6, 0.7$.
}
\end{figure}

\begin{figure}[h]
	\centering 
	\includegraphics[width=.45\textwidth]{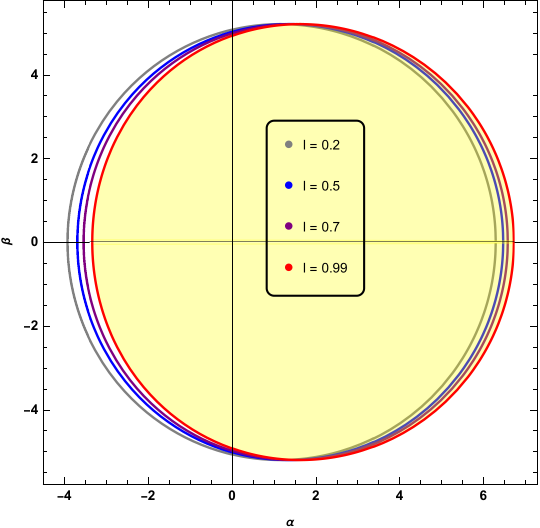}
	\hfill
	\includegraphics[width=.45\textwidth]{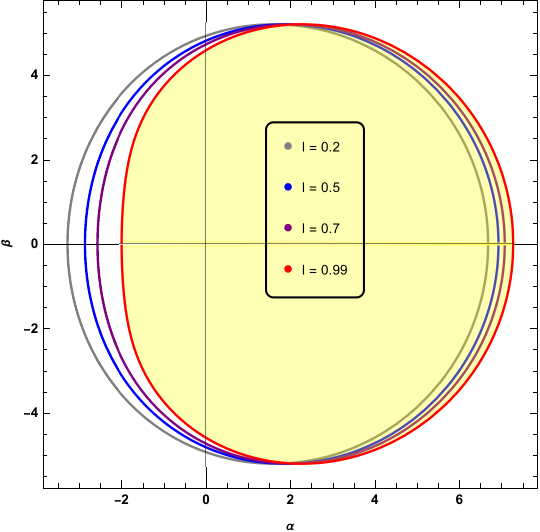}
	\hfill
	\includegraphics[width=.45\textwidth]{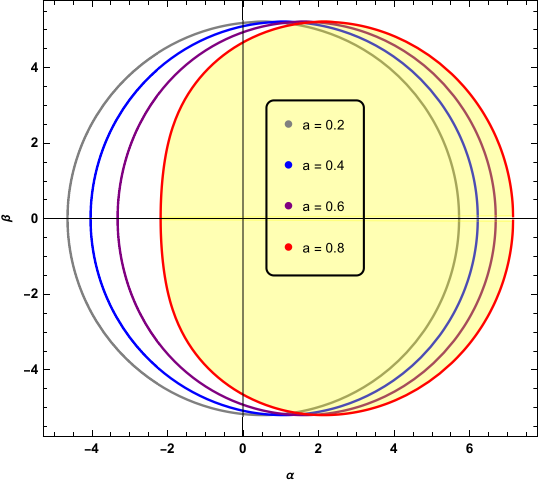}
	\hfill
	\includegraphics[width=.45\textwidth]{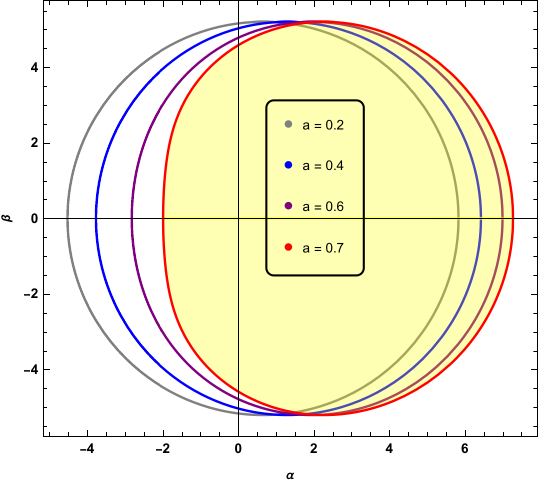}
	\caption{\label{fig:4} Variation in the shadow shape of the bumblebee BH M87 in the absence of DM distribution taking $\tilde{r}_{0} \simeq 16.8 \,\text{Mpc}$ with setting $\text{kpc}=1$.
The upper left panel corresponds to $a=0.5$ and $\ell=0.2, 0.5, 0.7, 0.99$, the upper right panel corresponds to $a=0.7$ and $\ell=0.2, 0.5, 0.7, 0.99$, the lower left panel corresponds to $\ell=0.5$ and $a=0.2, 0.4, 0.6, 0.8$, and the lower right panel corresponds to $\ell=0.99$ and $a=0.2, 0.4, 0.6, 0.7$.	
}
\end{figure}

\begin{figure}[h]
	\centering 
	\includegraphics[width=.45\textwidth]{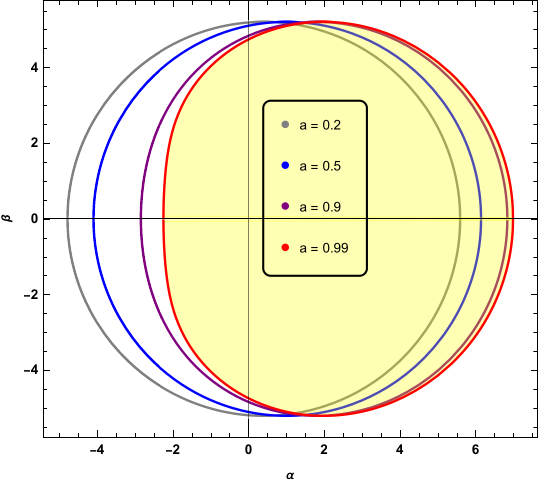}
	\caption{\label{fig:5} 
Variation in the shadow shape of the BH M87 in the absence of DM distribution and spontaneous breaking of Lorentz symmetry in terms of four different values of $a$.
}
\end{figure}

In Figs. \ref{fig:3} , \ref{fig:4} and \ref{fig:5}, we show the various shapes of the shadow by plotting $\beta$ versus $\alpha$, where, in Fig. \ref{fig:3}, the power index of the DMS is given by $\gamma_{\mathrm{SP}}=7/3$, corresponding to an initial DM distribution that follows the NFW profile, as prescribed by the CDM model.

The SMBH at the center of M87 has recently been observed by the EHT collaboration \cite{EventHorizon,EventHorizonL4,EventHorizonL6}, providing a measurement of its size and outline through the observed shadow. As more observational data are accumulated by EHT, it is becoming increasingly feasible to use the BH shadow to examine the distribution and characteristics of matter in the vicinity of the BH. While the impact of the DM halo on the BH shadow (and its metric) remains minimal when compared to the precision of EHT observations, this would not be the case if there were a DMS present \cite{XuJCAP2021}. 
According to Nampalliwar et al. \cite{NampalliwarAJ2021}, a significant variation in the size of the BH shadow can be observed for a range of DMS densities, exceeding the $17\%$ uncertainty reported by the EHT collaboration \cite{EventHorizonL6}. However, this change occurs only at the highest DM density levels. 
As reported by the EHT collaboration \cite{EventHorizon,EventHorizonL6}, the angular size of the M87 shadow is $\theta_{s}= (42 \pm 3)\text{$\mu$as}$, and the distance between the observer and the BH M87 is $D_{0} = 16.8\, \text{Mpc}$. Additionally, based on the association of the crescent feature in the M87 image with the emission surrounding the BH shadow and the knowledge of the distance to M87, the EHT collaboration estimated the mass of the M87 BH to be $M_{\mathrm{BH}} = (6.5 \pm 0.9) \times 10^{9}\, M_{\odot}$ \cite{KhodadiJCAP2020}.
To estimate the impact of the DMS on observable quantities like the shadow radius/angular diameter of the M87 BH, we may consider a specific scenario in which the core radius is significantly small, for instance, when $r_{0}\sim1.71\, \text{kpc}$ and $\rho_{0}r_{0}^{3} \sim 0.005\, M_{\mathrm{BH}}$ [1].  It is worth noticing that as $\rho_{0}r_{0}^{3}$ increases, the radius of the shadow also increases \cite{HouJCAP2018}. Furthermore, if we decrease the core radius while keeping $\rho_{0}$ constant, the shadow images decrease in size.
Therefore, investigating the effect of a DMS on the shadow of bumblebee BHs is meaningful. We have to note that the impact of DM distribution on the shadow radius/angular diameter is more pronounced in the spike region (i.e., where the DM density is high) when utilizing the Bumblebee Gravity model.

As it can be seen in Fig. \ref{fig:3}, the shadow shapes of the bumblebee BH in M87 surrounded by DMS are significantly affected by the values of the parameters $a$ and $\ell$. The upper panel of the figure reveals that the shadow becomes increasingly distorted as $a$ increases to $1$, with $\ell$ being fixed. The upper left panel displays the variation in the shape of the BH shadow in the absence of the LV parameter, while the upper right panel presents this variation at $\ell=0.99$ for the bumblebee BH shadow case. This implies that the effect of the $a$ and $\ell$ parameters on the shadow shape becomes more significant as $\ell$ becomes larger and closer to $1$.

Then, we can investigate the variation in the shape of the bumblebee BH shadow by fixing the spinning parameter $a$ and considering four different values of $\ell$. As shown in the lower panel of the figure, when $a$ is fixed at $0.2$ (left panel), the shadow shape is relatively insensitive to changes in $\ell$. However, for a faster-rotating BH with $a=0.99$ (right panel), the shadow shape changes more significantly with increasing $\ell$. Thus, we can conclude that bumblebee BHs with higher spin exhibit a greater sensitivity to variations in the $\ell$ parameter, indicating that the shadow shape is more affected by changes in this parameter.

Fig. \ref{fig:4} allows us to investigate the impact of the $a$ and $\ell$ parameters on the bumblebee BH shadow in the absence of DM distribution. In the upper left panel, where we fix the $a$ parameter at $0.5$, we observe that the shadow disk remains relatively undeformed for four different values of $\ell$, despite its tendency to stretch to the right side of the coordinate frame as $\ell$ approaches $1$. In contrast, in the upper right panel, where we fix the $a$ parameter at $0.7$, we observe significant deformation in the shadow as $\ell$ increases to $1$. The lower panel further illustrates this effect by fixing $\ell$ at $0.5$ and $0.99$, respectively, from left to right. As the value of the spin parameter increases, the deformation of the shadow becomes increasingly apparent.

Fig. \ref{fig:5} illustrates the impact of the spinning parameter on the shadow of a standard Kerr vacuum BH when there is no DM distribution or spontaneous breaking of Lorentz symmetry affecting the space-time. As the spinning parameter values increase, the shape of the shadow of this Kerr BH becomes distorted.

\begin{figure}[h]
\centering 
\includegraphics[width=.45\textwidth]{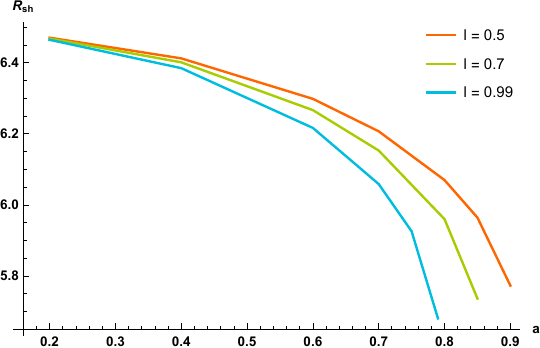}
\hfill
\includegraphics[width=.45\textwidth]{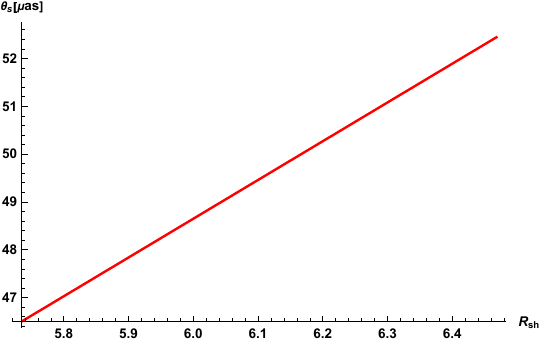}
\hfill
\includegraphics[width=.45\textwidth]{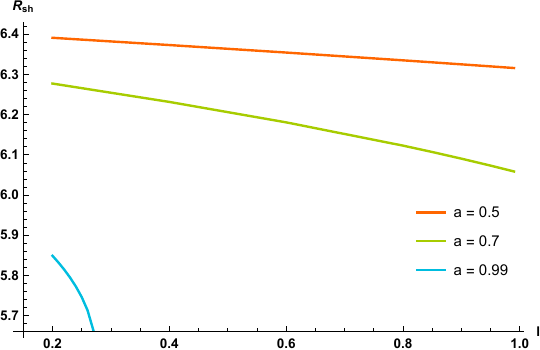}
\hfill
\includegraphics[width=.45\textwidth]{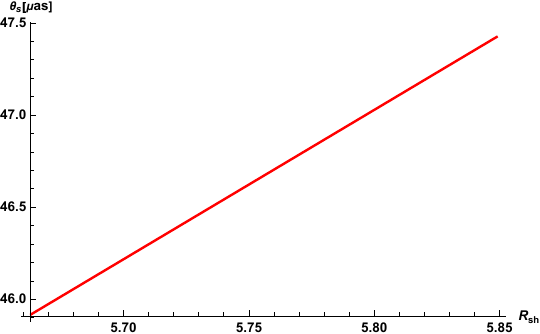}
\caption{\label{fig:6} Variation in the shadow radius $R_{\mathrm{sh}}$ and the angular diameter $\theta_{s}$  of the bumblebee BH M87 surrounded by DM distribution using $\mathcal{N}_{\gamma}=0.1$, $\gamma=1$, $\gamma_{\mathrm{SP}}=7/3$, 
$r_{0}\simeq1.71\,\text{kpc}$, $\rho_{0}\simeq 6.9\times10^{6}\,(M_{\odot}/\text{kpc}^{3})$,  $M_{\mathrm{BH}} \simeq 6.5\times 10^{9}M_{\odot}$
$R_{\mathrm{S}} \simeq 6.31 \times 10^{-7}\,\text{kpc}$, $R_{\mathrm{SP}} \simeq 2.347\, \text{kpc}$,  $\rho_{\mathrm{R}}\simeq 0.0008\, (M_{\mathrm{BH}}/\text{kpc}^{3})$ and $\tilde{r}_{0} = D_{0} \simeq 16.8 \,\text{Mpc}$
for different values of $a$ and $\ell$. We take $M_\mathrm{BH} = 1$  in units of the BH M87 mass.	    
Here we have set $\text{kpc}=1$.  
The upper left panel corresponds to the variation of $R_{\mathrm{sh}}$ with respect to $a$ for the  varying parameter $\ell$, the upper right panel corresponds to $\theta_{s}$ with respect to $R_{\mathrm{sh}}$ in terms of $\ell = 0.7$ and $a\in(0.2,0.85)$,
the lower left panel corresponds to the variation of $R_{\mathrm{sh}}$ with respect to $\ell$ for the  varying parameter $a$, and the lower right panel corresponds to $\theta_{s}$ with respect to $R_{\mathrm{sh}}$ in terms of $a = 0.99$ and $\ell \in(0.2,0.99)$.
}
\end{figure}

\begin{figure}[h]
	\centering 
	\includegraphics[width=.45\textwidth]{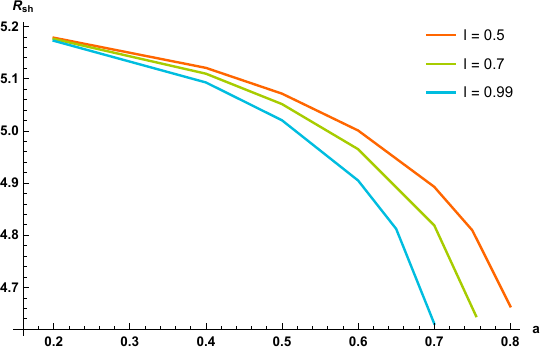}
	\hfill
	\includegraphics[width=.45\textwidth]{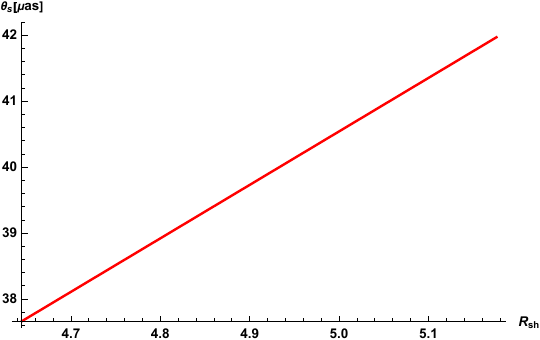}
	\hfill
	\includegraphics[width=.45\textwidth]{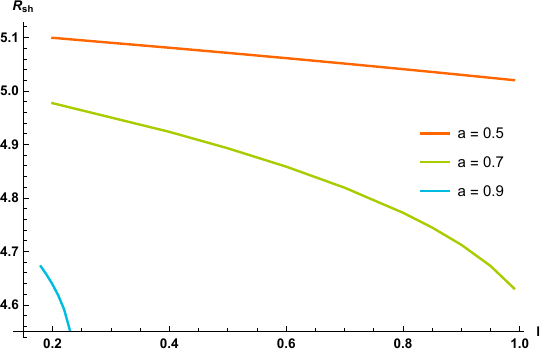}
	\hfill
	\includegraphics[width=.45\textwidth]{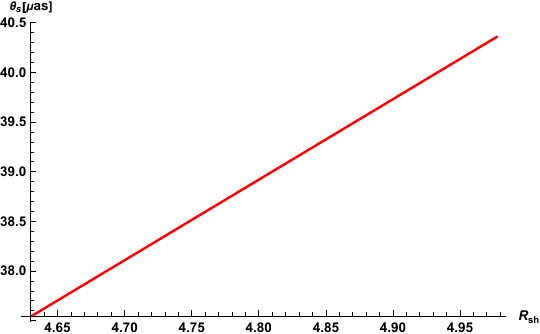}
	\caption{\label{fig:7} Variation in the shadow radius $R_{\mathrm{sh}}$ and the angular diameter $\theta_{s}$ of the bumblebee BH M87 in the absence of DM distribution taking $\tilde{r}_{0} =D_{0} \simeq 16.8 \,\text{Mpc}$ with setting $\text{kpc}=1$.
	The upper left panel corresponds to the variation of $R_{\mathrm{sh}}$ with respect to $a$ for the varying parameter $\ell$, the upper right panel corresponds to $\theta_{s}$ with respect to $R_{\mathrm{sh}}$ in terms of $\ell = 0.7$ and $a\in(0.2,0.75)$,
	the lower left panel corresponds to the variation of $R_{\mathrm{sh}}$ with respect to $\ell$ for the varying parameter $a$, and the lower right panel corresponds to $\theta_{s}$ with respect to $R_{\mathrm{sh}}$ in terms of $a = 0.7$ and $\ell \in(0.2,0.99)$.
	}
\end{figure}

In this way, we are interested in further exploring this scenario by computing the angular diameter of the M87 BH shadow. To achieve this, we will employ the observable quantity $R_{\mathrm{sh}}$ and calculate $\theta_{s}$ according to the formula $\theta_{s}=2R_{\mathrm{sh}}M_{\mathrm{BH}}/D_{0}$ \cite{NampalliwarAJ2021}.
Alternatively, we can rewrite this formula as 
\begin{equation}
\theta_{s}=2\times9.87098\times 10^{-6}R_{\mathrm{sh}}\frac{M_{\mathrm{BH}}}{M_{\odot}}\frac{1\,\text{kpc}}{D_{0}}\,\mu as .
\end{equation}
In Fig. \ref{fig:6}, we show how the shadow size and angular diameter change for different values of $a$ and $\ell$ under the effect of the nontrivial topology of the surrounding DMS. For values of $(a,b)$ between 0 and 1, the diameter of the shadow changes in the interval $2R_{\mathrm{sh}}^{\mathrm{DMS}}\in(11,13)$ in the presence of DMS. Additionally, when the DM effects are stronger, such as when the DM density is high, the angular diameter shifts. In Fig. \ref{fig:7}, that is for the case without DMS or when the effect of DM is negligible, the diameter of the shadow changes in the interval $2R_{\mathrm{sh}}^{\mathrm{non-DMS}}\in(9,10.5)$ for the same range of $(a,b)$. It should be noted that the reported value of $42 \,\mu\text{as}$ for the angular size of the shadow of M87 remains consistent in the absence of DMS.

\section{Discussion and Conclusions
	\label{Conc}
}
The recent discovery of the BH at the center of the Virgo A galaxy, M87, has led us to develop a new model in the framework of the Bumblebee Gravity. We propose that the central regions, including the BH itself, are surrounded by a power-law density profile DMS. To achieve this, we constructed a bumblebee BH solution surrounded by DM distribution, in the spike region and beyond, and selected a power-law density profile suitable for this type of galaxy. Using the density model, we first constructed the space-time metric for a non-rotating Schwarzschild bumblebee BH and then extended it to the rotating case for a more realistic model. We utilized both Newton approximation and perturbation approximation  to solve the Einstein field equations following the Xu et al. method.
We then investigated the impact of the DMS profile and the LV parameter, which account for the effects of Bumblebee Gravity on the various features of the Kerr bumblebee BH, including  the horizons,  the static limit surfaces,  the ergo-region, and the shadow images. 

Within this framework, different power-law indices of $\gamma_{\mathrm{SP}}$ correspond to distinct DM models at large distances outside of the DMS radius (i.e., $r \gg R_{\mathrm{SP}}$). By matching the DMS metric with the outside metric obtained via the CDM or TF model, we can examine different DM models. Our particular focus lay on the CDM model, which is characterized by a $\gamma_{\mathrm{SP}}$ value of $7/3$.

Clearly, the impact of the DM distribution on the shadow and other observable features depends  on the mass distribution in the immediate vicinity of the BH at the center of the galaxy. For this reason, we have disregarded the DM distribution in the region $r\gg R_{\mathrm{SP}}$, which is far away from the spike region.

Our investigation of the Kerr bumblebee BH, modeled after the observed BH in M87, revealed that the inclusion of a DMS in its vicinity leads to significant effects in various features, including the event horizon, outer static limit surface, photon orbit radii, shadow radius and angular diameter. Furthermore, we found that regardless of the presence or absence of a DMS, an increase in the value of $\ell$ for a fixed spin parameter, $a$, results in an increase of  both the Cauchy horizon and inner static limit surface, while  the event horizon and outer static limit surface decrease. Notably, our analysis showed that the size of the horizons and static limit surfaces is more strongly affected by the LV parameter $\ell$ when the spin parameter is fixed at a larger value, corresponding to a faster BH. We also observed that increasing the $\ell$ parameter leads to a larger ergo-region area. 

Our analysis points out that the effect of the presence of the DMS on the bumblebee BH shadow observables, such as the shadow radius and angular diameter, is more significant compared to the effect of the DM halo on the BH shadow observables. It leads to a shift in the angular size of the M87 shadow. As the DM density increases, such as in the case of the DMS density, it is expected that the shadow radius and angular diameter will  significantly grow,  due to the increase in $\rho_{0}r_{0}^{3}$. The effect of the DM halo on the BH shadow is still small compared to the observational accuracy of EHT. However, if there is a DMS, the results are expected to be different.

Finally, our study demonstrated that as the values of $a$ and $\ell$ increase, the size of the shadow radius and angular diameter of the bumblebee BH also decrease, even for a fixed value of $\rho_{0}r_{0}^{3}$. This suggests that the effect of the non-trivial topology of the DMS on the BH shadow observables becomes more significant as $a$ and $\ell$ increase.
Furthermore, we also observed that the LV parameter $\ell$ has a crucial impact on the bumblebee BH shadow and its observables, particularly when the spin parameter $a$ is close to one.
Specifically, forthcoming observations are expected to fix better the range of parameters $a$ and $\ell$.
Overall, the present model may shed new light on the nature of the interplay between DMSs and bumblebee BHs. 

\acknowledgments{
The authors thank  the Referee for the careful review of the manuscript and valuable suggestions.
This paper is based upon work from COST Action CA21136 {\it Addressing observational tensions in cosmology with systematics and fundamental physics} (CosmoVerse) supported by COST (European Cooperation in Science and Technology). SC  acknowledges the  {\it Istituto Nazionale di Fisica Nucleare}  ({\it iniziative specifiche} QGSKY and MOONLIGHT2).} DFM thanks the Research Council of Norway for their support and the resources provided by UNINETT Sigma2 -- the National Infrastructure for High Performance Computing and Data Storage in Norway.

\end{document}